\documentclass[journal]{IEEEtran}

\usepackage{threeparttable}
\usepackage{graphicx}
\usepackage{amsmath}
\usepackage{cite}
\usepackage{array}
\usepackage{algorithm}  
\usepackage{algpseudocode}  
\usepackage{amsmath}
\usepackage{setspace}

\begin{document}

\title{A Reconfigurable Convolution-in-Pixel CMOS Image Sensor Architecture}

\author{Ruibing~Song,~\IEEEmembership{Student Member,~IEEE,}
        Kejie~Huang,~\IEEEmembership{Senior Member,~IEEE,}
        Zongsheng~Wang,~\IEEEmembership{Student Member,~IEEE,}
        and~Haibin~Shen
\thanks{K.Huang and H.Shen are with the College of Information Science \& Electronic Engineering, Zhejiang University, 38 Zheda Road, Hangzhou, China, 310027, and also with Zhejiang Lab, Building 10, China Artificial Intelligence Town, 1818 Wenyi West Road, Hangzhou City, Zhejiang Province, China,
email: huangkejie@zju.edu.cn; shen\_hb@zju.edu.cn}
\thanks{R.Song and Z.Wang are with the College of Information Science \& Electronic Engineering, Zhejiang University, 38 Zheda Road, Hangzhou, China, 310027, email: songruibing@zju.edu.cn; wangzongsheng@zju.edu.cn}}

\markboth{IEEE TRANSACTIONS ON CIRCUITS AND SYSTEMS FOR VIDEO TECHNOLOGY,~Vol.~X, No.~X, September~2021}%
{Song \MakeLowercase{\textit{et al.}}: A Reconfigurable Convolution-in-Pixel CMOS Image Sensor Architecture}

\maketitle

\begin{abstract}
The separation of the data capture and analysis in modern vision systems has led to a massive amount of data transfer between the end devices and cloud computers, resulting in long latency, slow response, and high power consumption. Efficient hardware architectures are under focused development to enable Artificial Intelligence (AI) at the resource-limited end sensing devices. One of the most promising solutions is to enable Processing-in-Pixel (PIP) scheme. However, the conventional schemes suffer from the low fill-factor issue. This paper proposes a PIP based CMOS sensor architecture, which allows convolution operation before the column readout circuit to significantly improve the image reading speed with much lower power consumption. The simulation results show that the proposed architecture could support the computing efficiency up to 11.65 TOPS/W at the 8-bit weight configuration, which is three times as high as the conventional schemes. The transistors required for each pixel are only 2.5T, significantly improving the fill-factor.
\end{abstract}


\IEEEpeerreviewmaketitle

\section{Introduction}

\IEEEPARstart{W}{ith} the rapid development of image sensors and computer vision, the machines now can ``see” and ``understand” the visual world. Among various artificial neural networks, Convolutional Neural Network (CNN) has become dominant in various computer vision tasks and is attracting interest across a variety of domains, including object detection \cite{movingdetect}, face recognition \cite{facerecognition}, video compression \cite{videowulirong}, motion transfer \cite{motionweidongxu}, etc.

Although CNN has significantly improved visual systems' performance, they consume numerous operations and huge storage space, making it difficult for end devices to complete the computation independently. Therefore, data capture and analysis are separately carried out in modern visual systems by sensing devices and cloud computers, respectively. A tremendous amount of data transfer leads to a long delay, slow response, and high power consumption \cite{naturepis}. Moreover, in many vision applications, the systems have to work continuously for monitoring or anomaly detection, i.e., surveillance cameras. The low information density has seriously wasted communication bandwidth, data storage, and computing resource. 

To improve the efficiency of modern vision systems, researchers are focusing on reducing the readout power consumption or data density of sensors \cite{pwmJSSC,pwmTCASi,edsensor,ulvpwm,aoJSSC,EHCIS}. One of the most promising methods is to enable the computing at the sensing units, which has been intensively studied for with traditional algorithms. For example, \cite{onchipmod} computes the difference between the new pixel value and the previous value stored on the in-pixel analog memory to detect motion events, and locates the objects by on-chip computing circuits.
\cite{dpsmfe} adopts a Digital Pixel Sensor Array which has memory, readout circuits, and ADC in each pixel to perform row-parallel motion feature extraction. In-sensor computing with CNN has similar ideas. Computing CMOS Image Sensors (CIS) for CNN can be divided into three categories: (1) Processing-Near-Sensor (PNS), (2) Processing-In-Sensor (PIS), and (3) Processing-in-Pixel (PIP). The PNS architecture utilizes on-chip Deep Learning Accelerators (DLA) to shorten the physical distance between the processor and the image sensor \cite{3dstacked,neurosensor,pnsDLA}. The PIS architecture is proposed to reduce the data transfer distance, read operations, and analog-to-digital conversions. For example, Redeye performs several layers of CNN computing in CIS by additional analog arithmetic circuits before readout, saving 85\% energy due to the reduced read operations \cite{redeye}. However, it needs many analog capacitors for data storage, leading to a large area overhead and low computational efficiency. PIP is a fully integrated architecture to enable sensing and computing simultaneously. However, they need complicated pixel circuits, which lead to excessive area and power consumption \cite{2020Fully,2019IEDM}.

\begin{figure}[!t]
\centering
\includegraphics[width=0.5\textwidth]{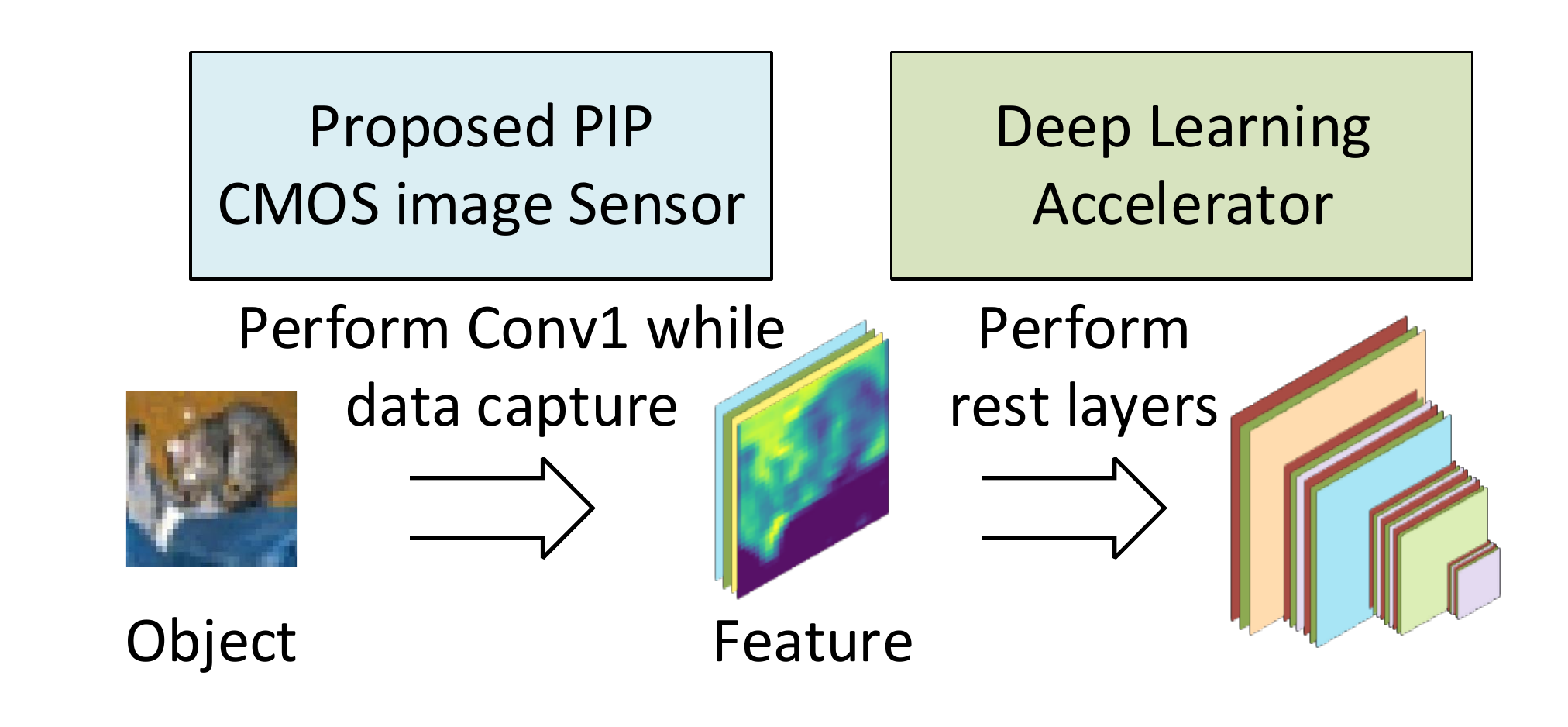}
\caption{The first layer of CNN is performed by PIP CMOS image sensor and the rest is done with DLA.}
\label{cnn-system}
\end{figure}

The convolution operation of the first layer is usually the performance bottleneck in various acceleration algorithms for the following reasons. The first reason is the limited number of input channels and parameters in the first layer, leading to the challenge in pruning or quantification. For example, the pruning in \cite{conv1} and quantification in \cite{BWN} leave the weight of the first layer unchanged, because it is very sensitive to the accuracy of the network. To support the entire network, DLA must be designed in accordance with the requirements of every layer, leading to the hardware complexity and the computing inefficiency. Furthermore, there are many differences between the structure of the first layer and the subsequent layers, so the computing resources cannot be effectively utilized. The PIP architecture is a possible solution that performs the computing of the first layer while sensing with high efficiency.

Based on the above reasons, we propose a novel PIP architecture as shown in Fig. \ref{cnn-system}, which moves the operations of the first layer to the sensor to improve the resource utilization and computing efficiency of the DLAs. 
The proposed PIP architecture can enable highly efficient convolution computation in pixels to improve both fill-factor and computing efficiency. The Multiply-accumulate (MAC) operation is achieved by Pulse Width Modulation (PWM) and pixel splicing. The entire pixel array allows massive parallel convolution operations, generating one complete output feature map in four steps when the stride is two, and the filter size is 3$\times$3$\times$3. Our proposed architecture could also support 60 frames and 128$\times$128 resolution when the output channel size is 64. 
Early works such as \cite{onchipmod} and \cite{dpsmfe} only support traditional algorithms and some PIP architectures such as \cite{2020TCASii} and \cite{nearsensor} only support Binary Neural Networks, while our proposed scheme can support various convolution kernels via the proposed “kernel splicing” method. The projected computational efficiency can be as high as 11.65 TOPS/W, which is three times higher than that in the reported literature \cite{3dstacked,2019ASSC,2020TCASii,nearsensor,2021pip}.

The rest of this paper is organized as follows: Section \uppercase\expandafter{\romannumeral2} presents the related works. Section \uppercase\expandafter{\romannumeral3} introduces the detailed design of our proposed scheme, including the overview architecture, the pixel circuit, the MAC operation, array convolution, and the implementation of other convolution kernel sizes. Section \uppercase\expandafter{\romannumeral4} analyzes the simulation results. Finally, the conclusion is drawn in Section \uppercase\expandafter{\romannumeral5}.

\section{Related Work}

\begin{figure}[!t]
\centering
\includegraphics[width=0.48\textwidth]{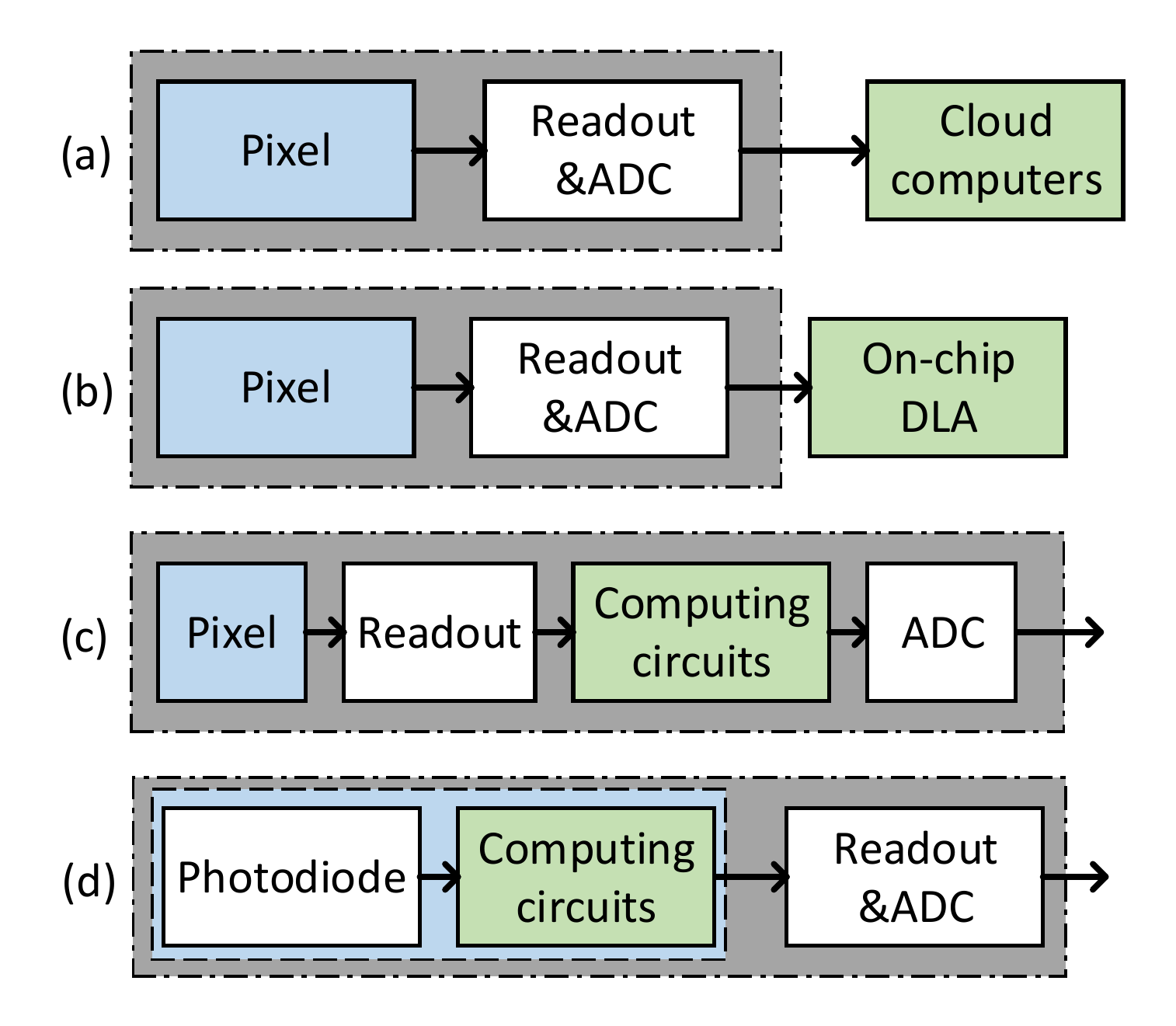}
\caption{Different architectures of visual systems. (a)Traditional architecture. (b)PNS architecture. (c)PIS architecture. (d)PIP architecture. Blue boxes represent the pixel, grey boxes mean the sensors, and green boxes show where the computing is conducted.}
\label{cis-architectures}
\end{figure}

Fig. \ref{cis-architectures} shows the block diagram of different architectures, including traditional architecture, PNS, PIS, and PIP. 


\textbf{Traditional architecture (Fig. \ref{cis-architectures}(a))}. As shown in \cite{CIS2005}, the data capture and data analysis in traditional schemes are usually done in the CIS and server, respectively. As a result, the majority of time and energy is consumed by data transfer. New visual systems should be developed to reduce response time and energy consumption.

\textbf{PNS architecture (Fig. \ref{cis-architectures}(b))}. \cite{2021sony} proposed a stacked, backside-illuminated CIS with a Digital Signal Processor (DSP) dedicated to CNN computation. In \cite{2017DATE}, the signals are quantized by the ramp Analog to Digital Converters (ADCs) and then computed by the on-chip stochastic-binary convolutional neural network processor. Compared with the traditional architecture shown in Fig. \ref{cis-architectures}(a), PNS architectures reduce the energy consumption of data movement. However, the energy consumed by the data readout and quantization is still not optimized.

\textbf{PIS architecture (Fig. \ref{cis-architectures}(c))}. In PIS architectures, the computing units are moved to the place before ADC to reduce quantization frequency. Unlike PNS, the computing in PIS is usually done in the analog domain. The CIS in \cite{nearsensor} can realize a maximum 5$\times$5 kernel-readout with a minimum of one stride step for convolution operations. Analog processing units directly process the readout signals without ADCs. In \cite{2019ICTA}, input images are captured in the current mode and transferred to the in-sensor analog computing circuit. However, both schemes only support binary neural networks. 

\textbf{PIP architecture (Fig. \ref{cis-architectures}(d))}. In PIP architectures, the computing units are integrated with the pixel array. \cite{2019ASSC} adopted a linear-response PWM pixel to provide a PWM signal for analog-domain convolution. The weights for multiplication are achieved by adjusting the current level and the integral time based on the pixel-signal pulse width. Meanwhile, accumulation is implemented by the current integration. However, the current level is generated by Digital-to-Analog Converter (DACs) according to the weights, which leads to extra power consumption. \cite{2020Fully} adopted a pixel processor array-based vision sensor called SCAMP-5. Each pixel contains 13 digital registers and seven analog memory registers to achieve various operations. \cite{2020TCASii} proposed a dual-mode PIS architecture called MACSen, which has many SRAM cells and computation cells in each unit of the array. Both schemes suffer from a large pixel area and a low fill-factor. 

New materials and devices are also developed for PIP architectures to improve the fill-factor. \cite{nature2d} proposed a $WSe_2$ two-dimensional (2D) material neural network image sensor, which uses a 2D semiconductor photodiode array to store the synaptic weights of the network. However, changing the photosensitivity of the photodiode may need additional DACs for each pixel to enable massive parallel computing.  

\begin{figure*}[!t]
\centering
\includegraphics[width=0.8\textwidth]{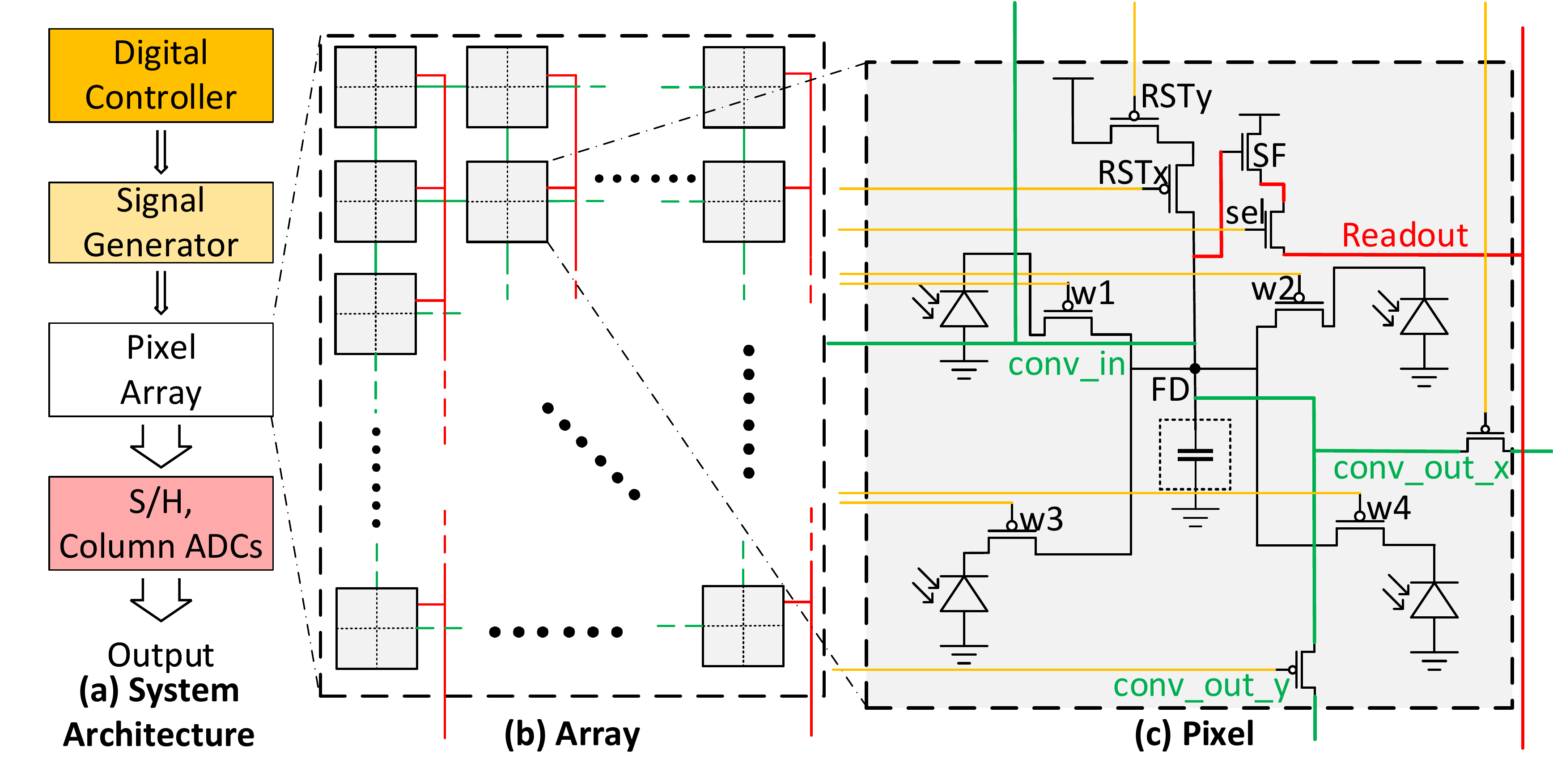}
\caption{The overview of the PIP architecture. (a) The system architecture, (b) The structure diagram of the pixel array, and (c) The pixel circuits. The red lines represent the column readout wires, the green lines represent the convlink wires between adjacent pixels, and the golden line shows the direction of the control signals.}
\label{proposed architecture}
\end{figure*}

\textbf{Mixed architecture}
It's usually difficult to conduct all computing tasks with only PIS or PIP architectures. Mixed schemes are thus proposed to achieve the entire neural network computing. In \cite{pisanalog}, an analog computing circuit is always-on to achieve face detection before ADCs. When faces are detected, the on-chip DLA performs the computing for face recognition in the digital domain, which can be described as a PIS + PNS scheme. \cite{2020crossbar} fabricated a sensor based on $WSe_2$/$h-BN$/$Al_2O_3$ van der Waals heterostructure to emulate the retinal function of simultaneously sensing and processing an image. An in-memory computing unit is added after the sensor to make up the PIP + PNS scheme.

\section{Proposed Architecture}
This section describes the detailed design of our proposed PIP architecture, as shown in Fig. \ref{proposed architecture}. The MAC operation is achieved by PWM and pixel splicing, significantly improving the pixel density and computing efficiency. The convolution operations are realized by reconfigurable switching at the array level, allowing massive parallel computing for high performance. The architecture supports the first layer of CNN, which has a low acceleration ratio due to small input channels and large feature maps. It is also very sensitive to pruning\cite{conv1} and quantization\cite{BWN}. The proposed CIS can work in the Traditional mode to output the raw image.

\subsection{Pixel Circuit and MAC Operation}
As shown in Fig. \ref{proposed architecture}(a), the digital circuit generates the control signals for the pixel array. The detailed designs of the pixel array and pixel circuit are illustrated in Fig. \ref{proposed architecture}(b) and (c), respectively. The convlink wires (shown as green lines) connect adjacent pixel units with splicing transistors in both row and column directions. Each column readout wire (shown as red lines) connects each column of pixel units to a Sample and Hold (S/H) circuit and column ADC, which are the same as traditional CIS. Fig. \ref{proposed architecture}(c) shows the circuit of a pixel unit containing four pixels. Four exposure control transistors $w_1-w_4$ are connected to a shared FD node. Two reset transistors $RST_x$ and $RST_y$, a source follower $SF$, and a row select transistor $sel$ are shared by four adjacent pixels representing RGGB channels. Two splicing transistors $conv\_out\_x$ and $conv\_out\_y$ control the connection of the convlink wires to the adjacent pixel units in the row direction and column direction, respectively. There are 10 transistors in total for four photodiodes in a pixel unit, and thus each pixel contains 2.5 transistors (2.5T) on average. 

\begin{figure}[!t]
\centering
\includegraphics[width=0.45\textwidth]{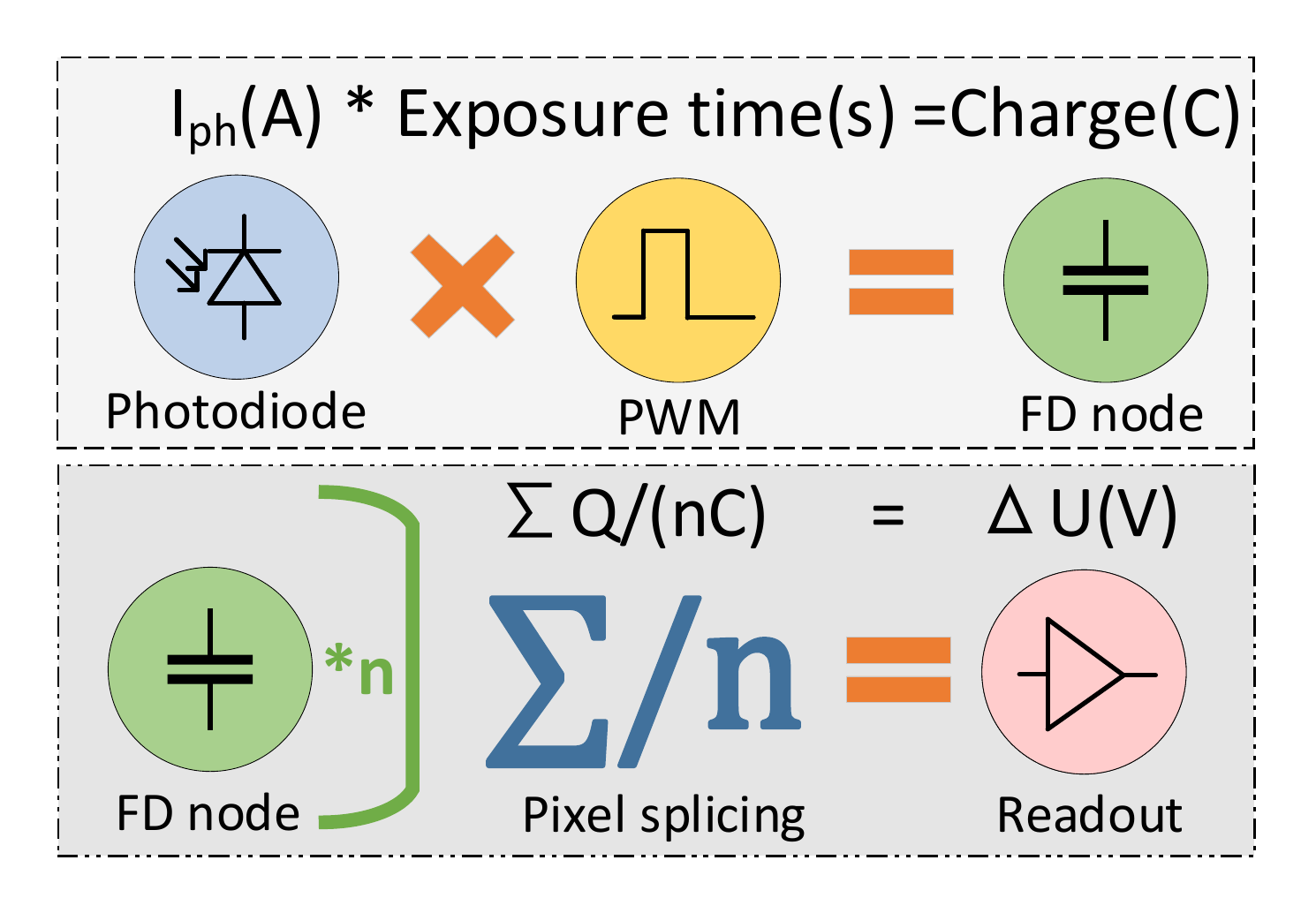}
\caption{The computing flow diagram of the proposed architecture.}
\label{computing flow}
\end{figure}

\begin{figure}[!t]
\centering
\includegraphics[width=0.5\textwidth]{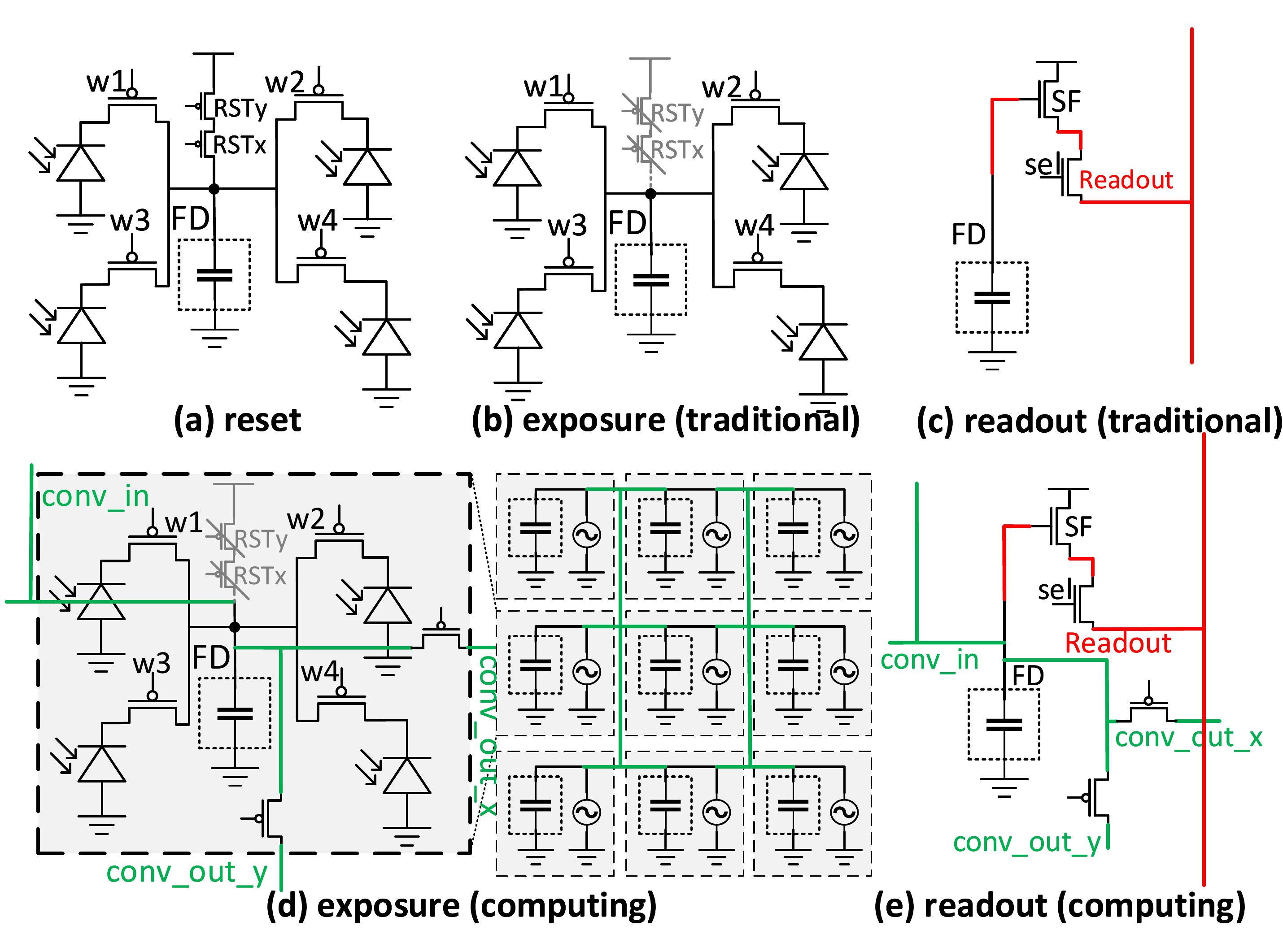}
\caption{Proposed pixel circuits in the (a) reset stage, (b) exposure stage in Traditional mode, (c) readout stage in Traditional mode, (d) exposure stage in Computing mode, (e) readout stage in Computing mode. The red lines represent the column readout wires and the green lines represent the convlink wires between adjacent pixels.}
\label{computing circuit}
\end{figure}

Fig. \ref{computing flow} shows the computing flow of the MAC operation with the proposed PIP architecture. The multiplication of photocurrent and weights is realized by controlling the exposure time of photodiodes in each pixel unit. The exposure time of photodiodes is modulated by the weight (8 bit) in the convolution kernel. Pixel splicing method links the FD nodes and averages the multiplication results, thus realizes the accumulation.

Fig. \ref{computing circuit} shows the simplified schematic of the proposed pixel in two different modes. Both Traditional and Computing modes include reset stage, exposure stage, and readout stage. The working mechanism of the Traditional mode is the same as the conventional 1.75T pixel array as shown in Fig.\ref{computing circuit} (a)-(c) and $RST_y$ is set to low to hand over the reset control to $RST_x$. The switch between Computing mode and Traditional mode can adopt an event-driven mechanism. When the target object is identified, the CIS can be switched to the Traditional mode to output the complete raw image. 

The timing diagram of the Computing mode is shown in Fig. \ref{pixel convolution}, which only contains four pixels for simplicity. A detailed description of each stage in exposure mode is provided as follows:

\subsubsection{}
In the RESET stage, as shown in Fig.\ref{computing circuit}(a), transistors $RST_x$, $RST_y$ and $w_1$-$w_4$ are turned on to reset the potential of the FD node and photodiodes to Vdd. For a Silicon PN photodiode, the rise time is about 5-10 ns \cite{photodiodebook}, which is shorter than our reset time (100 ns). Thus, the photodiode is saturated before the integration begins.

\subsubsection{}
The EXPOSURE stage is started after the RESET stage when RST is de-asserted, as shown in Fig.\ref{computing circuit}(d). In this stage, the control pulses of exposure signals $w_1$-$w_4$ are modulated by the weight. The exposure time $t_i$ is proportional to the weight value $w_i$. Assuming that the convolution kernel size is $r\times r$, one of the MAC operation results can be obtained by connecting $r^2$ adjacent pixel units with the convlink wires. We assume r = 3 as shown in Fig.\ref{computing circuit}(d). The circuit of each pixel unit can be simplified as the FD node connected in parallel with an AC current source. With the convlink wires, the nine FD nodes of the adjacent pixel units and nine AC current sources are connected in parallel. Since the photocurrent $I_i$ is unchanged in a short period, the charge $Q$ stored on each of these FD nodes can be expressed as

\begin{equation}
Q=CU_{rst}-\frac{\sum_{i=1}^{4r^2}(I_i+I_{dc})kw_i}{r^2}
\label{conv-basic}
\end{equation}
where $k$ is the exposure constant, adjusted by the software according to the external light intensity. $I_{dc}$ is the dark current and $C$ is the capacitance of the FD node. Thus, the charge $Q$ on each FD node represents the average of the products of the photocurrent $I_i$ and the corresponding weight value $w_i$ in the convolution kernel. The potential on FD nodes $U_{conv}$ can be expressed as

\begin{equation}
U_{conv}=U_{rst}-\frac{k}{r^2C}(\sum_{i=1}^{4r^2}[I_iw_i] - I_{dc}\sum_{i=1}^{4r^2}w_i)
\label{conv-uni}
\end{equation}

\begin{figure}[!t]
\centering
\includegraphics[width=0.48\textwidth]{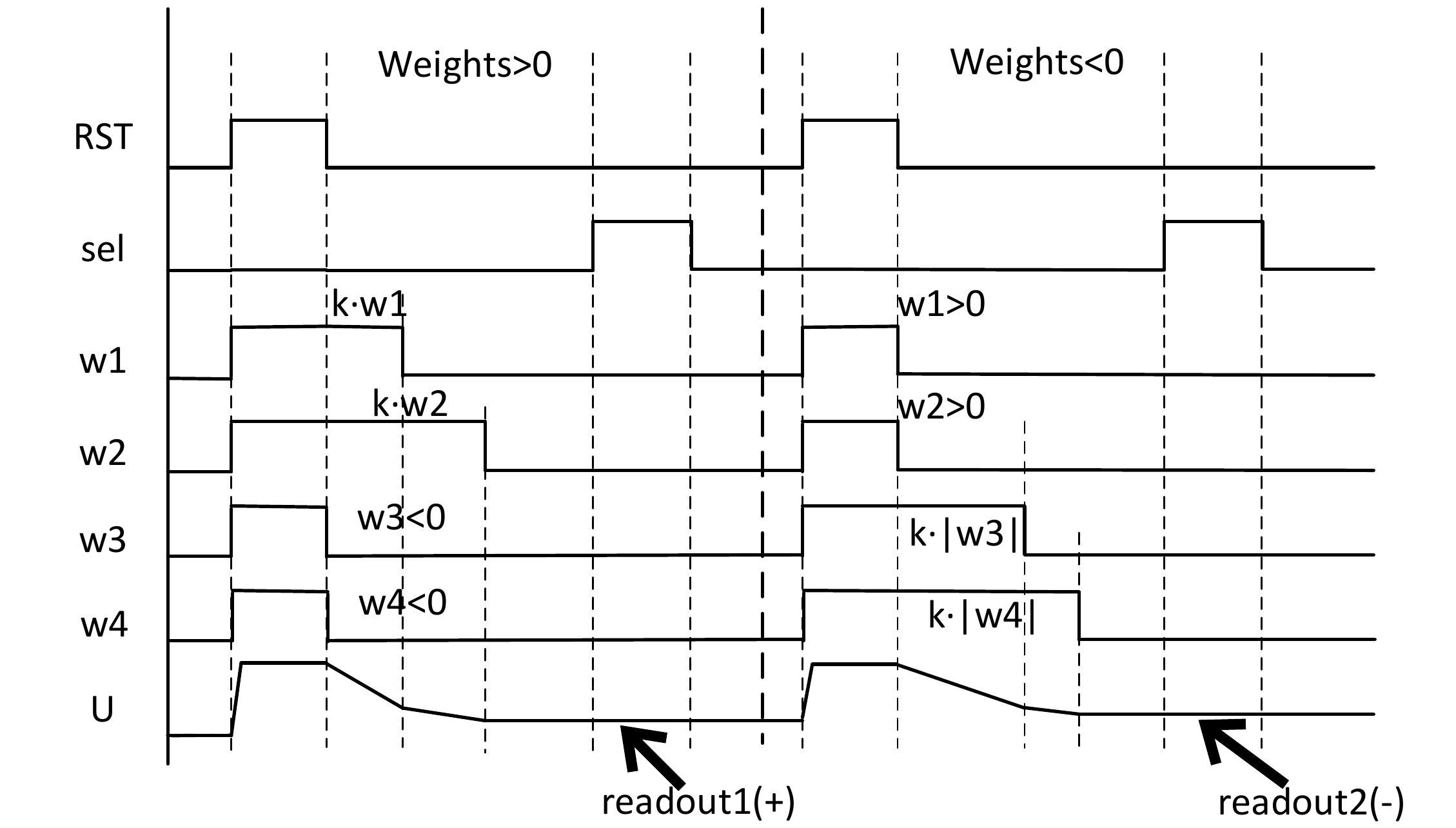}
\caption{The convolution sequence diagram of the pixel circuit.}
\label{pixel convolution}
\end{figure}

\begin{figure}[!t]
\centering
\includegraphics[width=0.5\textwidth]{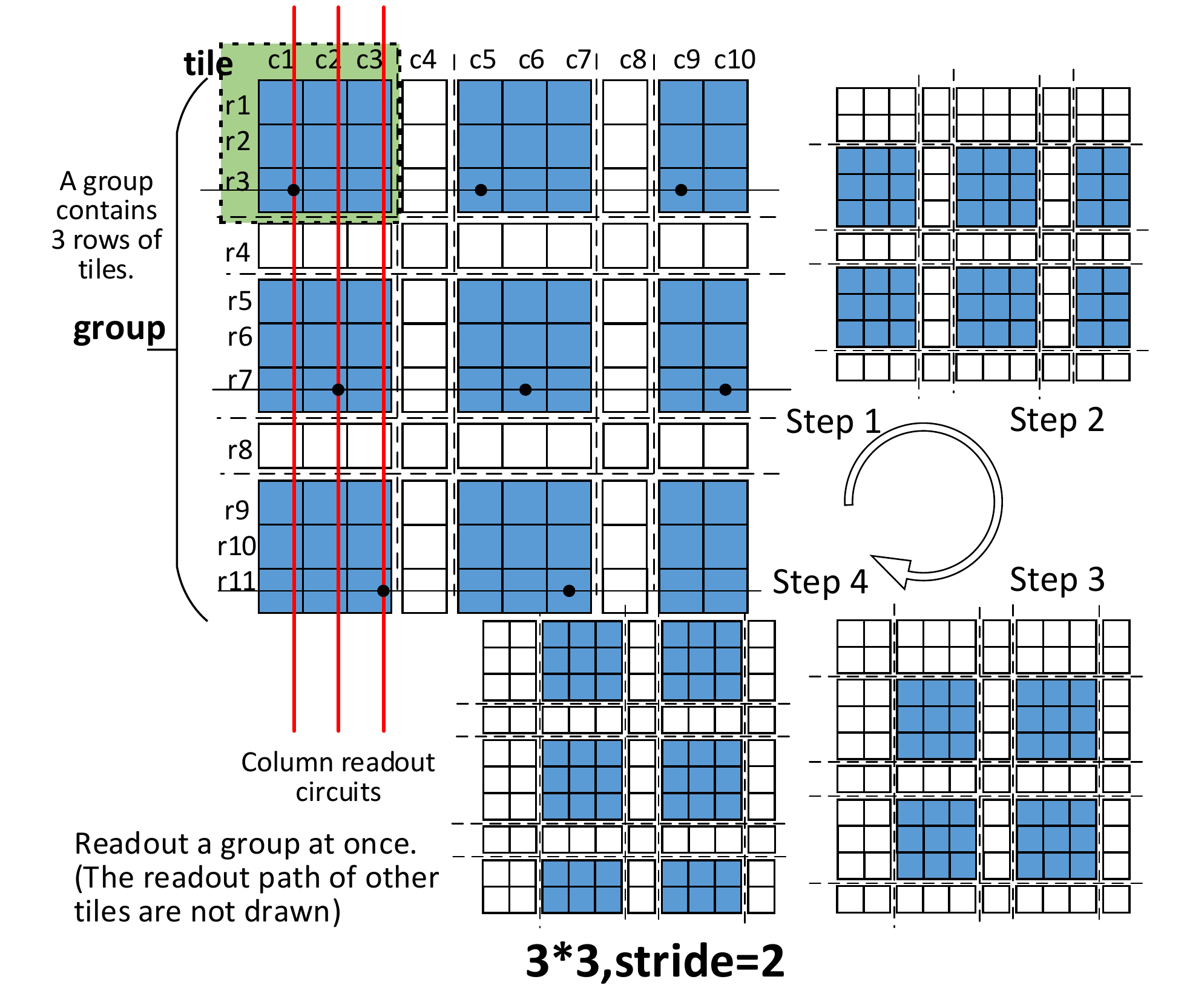}
\caption{The flow diagram of the array convolution operation. The convolution kernel size is 3x3, and the stride is 2. The blue squares represent the active pixels in each convolution step. The green area shows the definition of a tile as an example. The red line represents the column readout circuit.}
\label{conv33}
\end{figure}

\begin{figure*}[!t]
\centering
\includegraphics[width=0.8\textwidth]{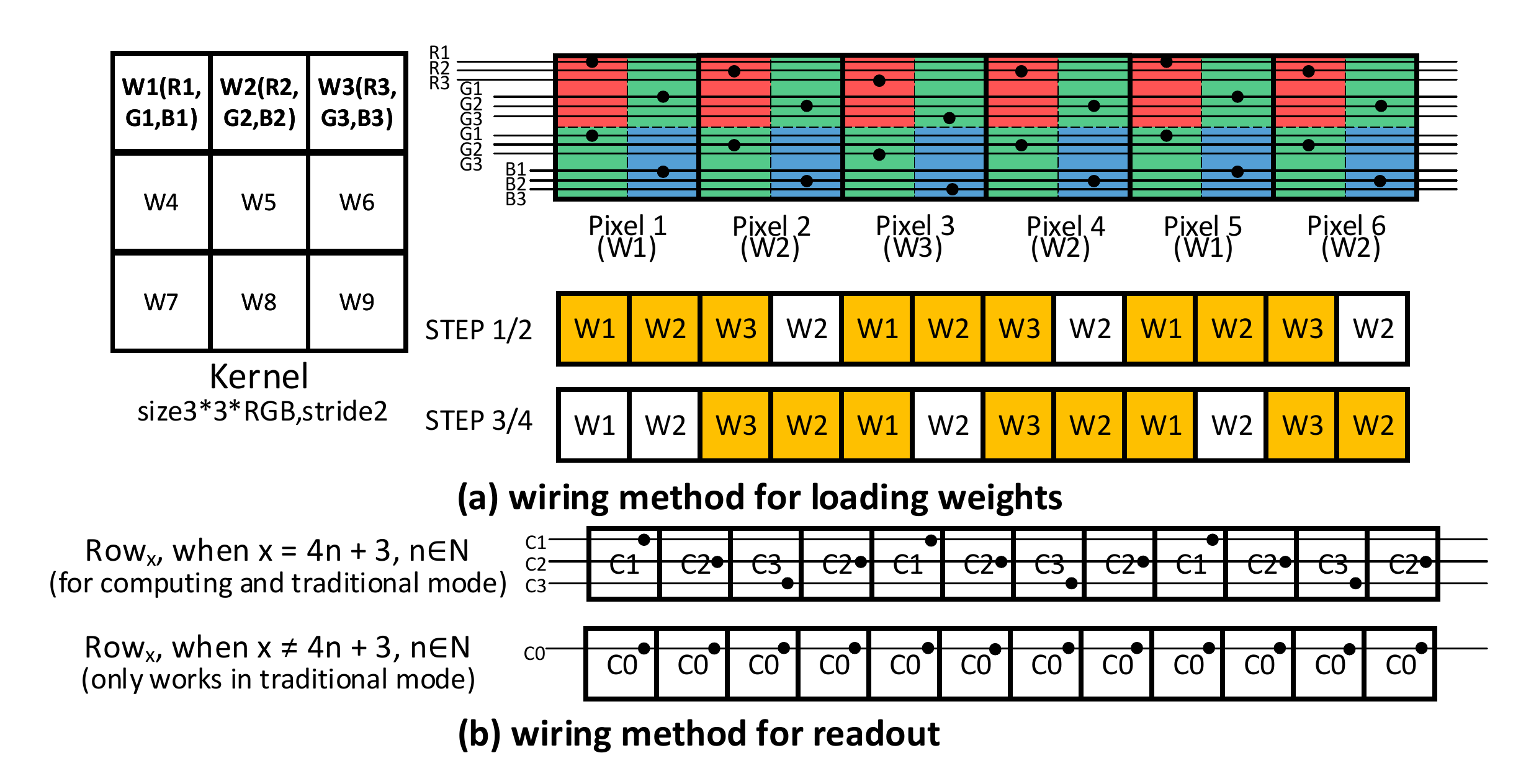}
\caption{(a) The wiring way of weight loading. The red, green, and blue squares show the filter RGGB pattern. The golden squares represent the active pixels in each step. (b) The wiring method for readout. The convolution kernel has a size of 3$\times$3, and the stride is 2.}
\label{wiring}
\end{figure*}

\subsubsection{}
After the exposure, it is the READOUT stage when $sel$ is asserted, as shown in Fig. \ref{computing circuit}(e). The weight precision of the convolution kernel used in the system is 8-bit. That is, the weight size of the convolution kernel ranges from -128 to +127. The positive and negative weights of the convolution kernel can be achieved by subtracting two consecutive exposures, as shown in Fig. \ref{pixel convolution}.  As $w_1$ and $w_2$ are positive, they are enabled in the first exposure period. The negative $w_3$ and $w_4$ are enabled in the second exposure period. The digital circuits subtract the two readout operations in Fig. \ref{pixel convolution} after the ADCs, which is expressed as

\begin{equation}
U=U^--U^+=\frac{k}{r^2C}(\sum I_iw_i^+ - \sum I_i|w_i^-| + I_{dc}\sum w_i)
\label{conv-pn}
\end{equation}
where $\frac{k}{r^2C}$ is a known constant, $\sum w_i$ is statistically close to zero and $I_{dc}$ is usually much smaller than $I_i$. The voltage $U_{conv}$ represents the sum of the $r^2$ multiplication results, thus achieving MAC operation in-pixel level.


\subsection{Convolution Operation in Array}

After introducing the basic idea of the MAC operation, this section gives a detailed introduction to the overall architecture of the system and the sliding convolution on the entire pixel array.

As can be seen from Fig. \ref{proposed architecture}, the most fundamental component of the pixel array is a pixel unit containing four pixels. Splicing transistors separate the adjacent pixel units. Each column of pixel units includes a column S/H circuit and a column ADC outside the array, which can read the convolution results and convert them into digital signals.

The flow of convolution operation in the array is shown in Fig. \ref{conv33}. In the following example, we assume that the convolution kernel size is 3$\times$3 and the stride is 2. In Fig. \ref{conv33}, each square represents a pixel unit. The 3*3 connected active pixel units are defined as a tile. The horizontal and vertical dash lines mean the break of convlink between the tiles. The entire array can be divided into several independent convolution tiles. The MAC operations are enabled simultaneously in all active tiles in each step. We defined three rows of tiles as a group. 

As stated in the previous section, the MAC operation can be achieved by connecting the convlink wires of all pixel units corresponding to a convolution kernel during computation. More MAC operations should be carried out simultaneously to maximize the parallel operation and computing throughput. Multiple simultaneous MAC operations regions must be non-overlapping, which is achieved by dividing the convolution procedures of the entire array into four steps, as shown in Fig. \ref{conv33}. The colored squares represent the active pixel units, and the uncolored squares represent pixel units not involved in each step. In such a scenario, all the convolution areas can be calculated and read out with only one exposure in one step. The active pixel units perform the MAC operations within the convolution kernel to get a quarter of the convolution result. After four steps of computing, a complete convolution operation is finished. As two exposures are required for the positive and negative weights of each step, eight exposure cycles are needed for each convolution operation.

The above convolution operation needs to plan the hardware wiring carefully. As shown in Fig. \ref{wiring}(a), when the convolution kernel size is 3$\times$3 and stride is 2, pixel units in the same row are connected to the wire in the following order: $W_1$, $W_2$, $W_3$, $W_2$, $W_1$, $W_2$, $W_3$... In this way, each tile in a step contains the same wire orders ``$W_1$, $W_2$, $W_3$" in the first and second steps and ``$W_3$, $W_2$, $W_1$" in the third and fourth steps. 
\begin{figure}[!t]
\centering
\includegraphics[width=0.48\textwidth]{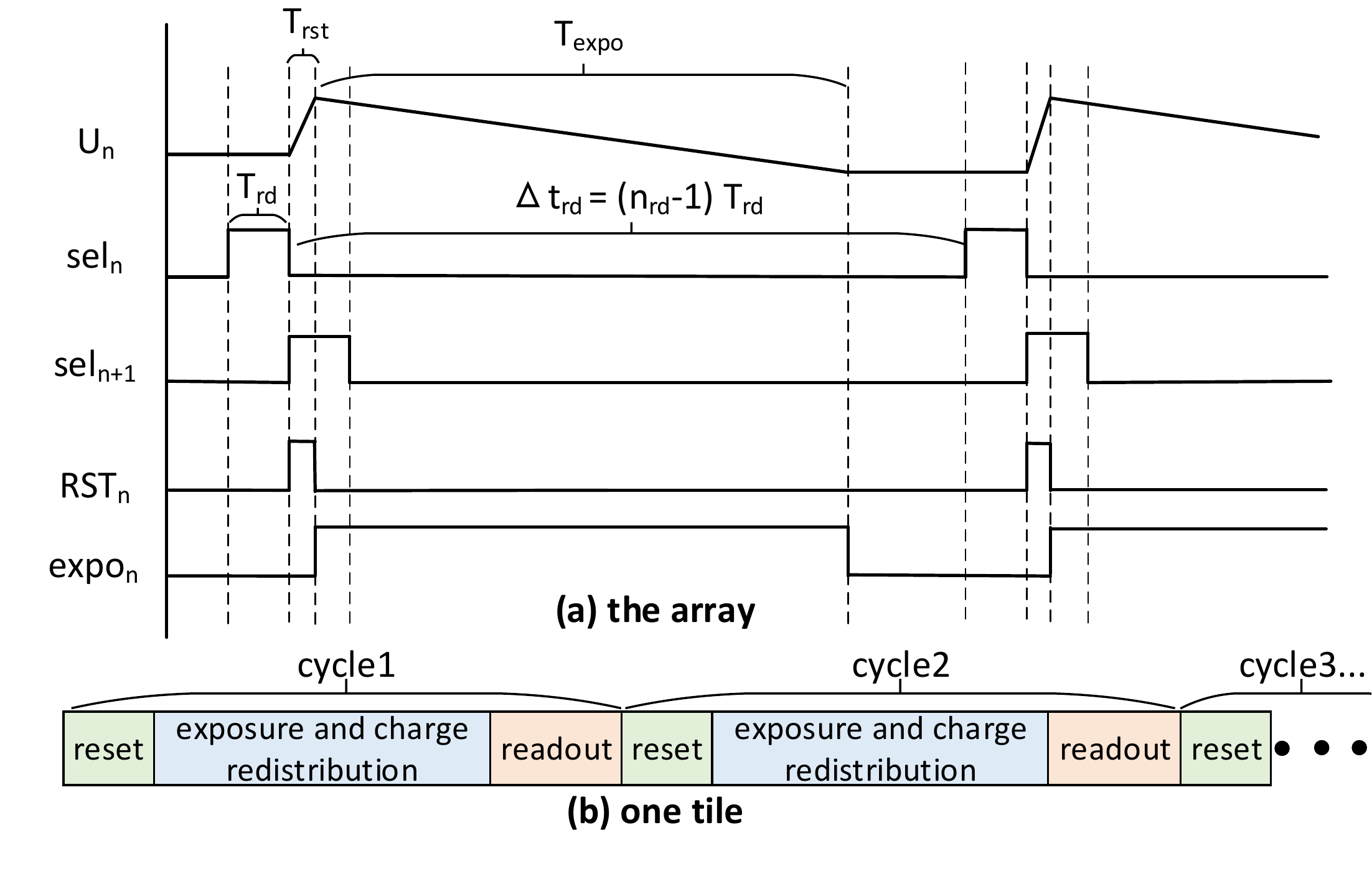}
\caption{The convolution operation sequence diagram of (a) the array and (b) one tile. $U$ represents the potential of the FD nodes in the chosen tiles, $RST$ represents the reset stage, $expo$ represents the exposure stage, and $rd$ represents the charge redistribution and readout stage.}
\label{array convolution}
\end{figure}

\begin{figure}[!t]
\centering
\includegraphics[width=0.48\textwidth]{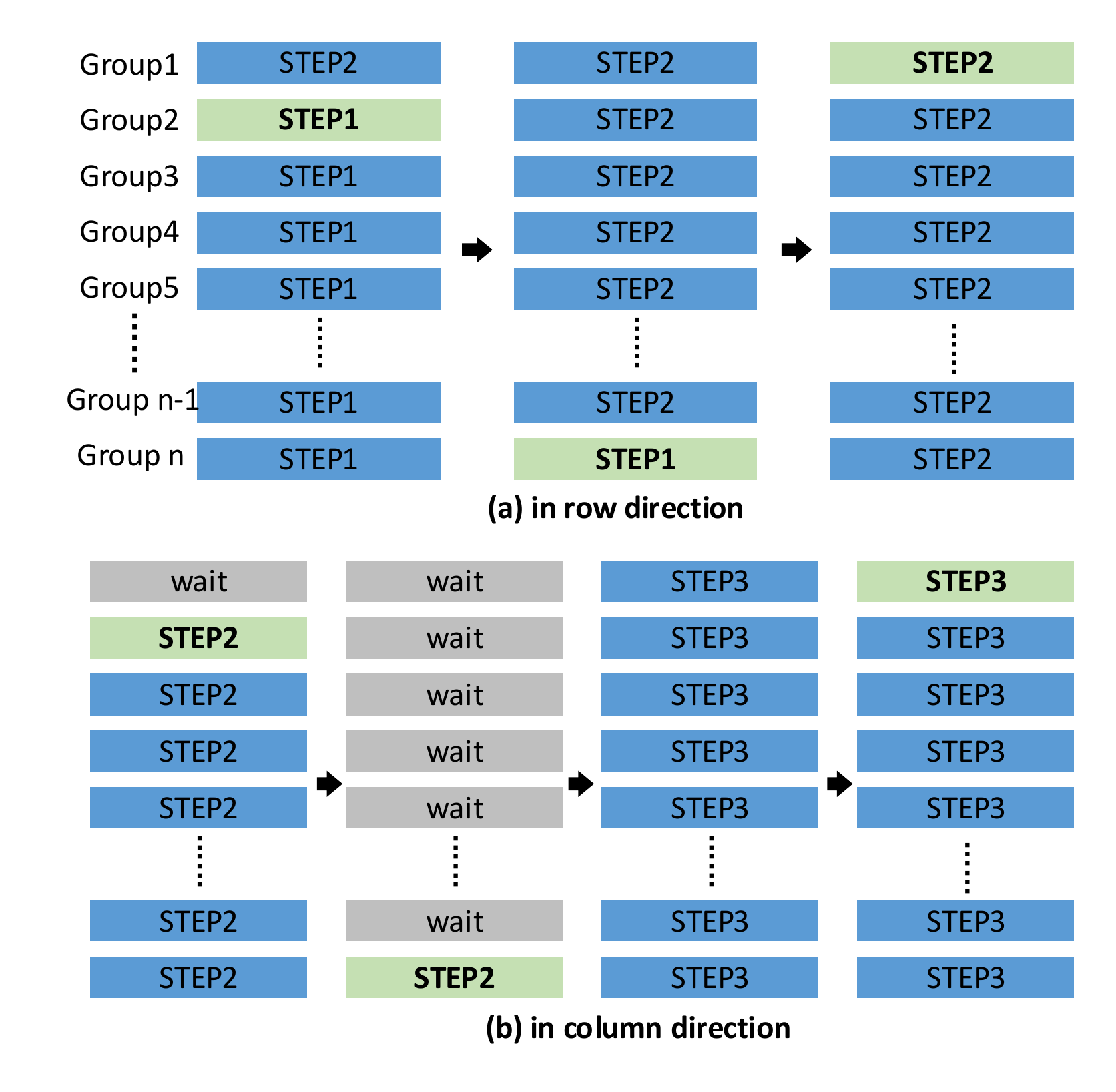}
\caption{The different processing sequence between steps when the control signals of splicing transistors are changed (a) in the row direction (such as step 1 to step 2) or (b) in the column direction (such as step 2 to step 3). The blue rectangles represent the groups are in exposure stage, the green rectangles represent the groups are in readout stage, and the grey rectangles mean the groups are waiting for next exposure stage.}
\label{cis-method}
\end{figure}
\begin{figure*}[!t]
\centering
\includegraphics[width=\textwidth]{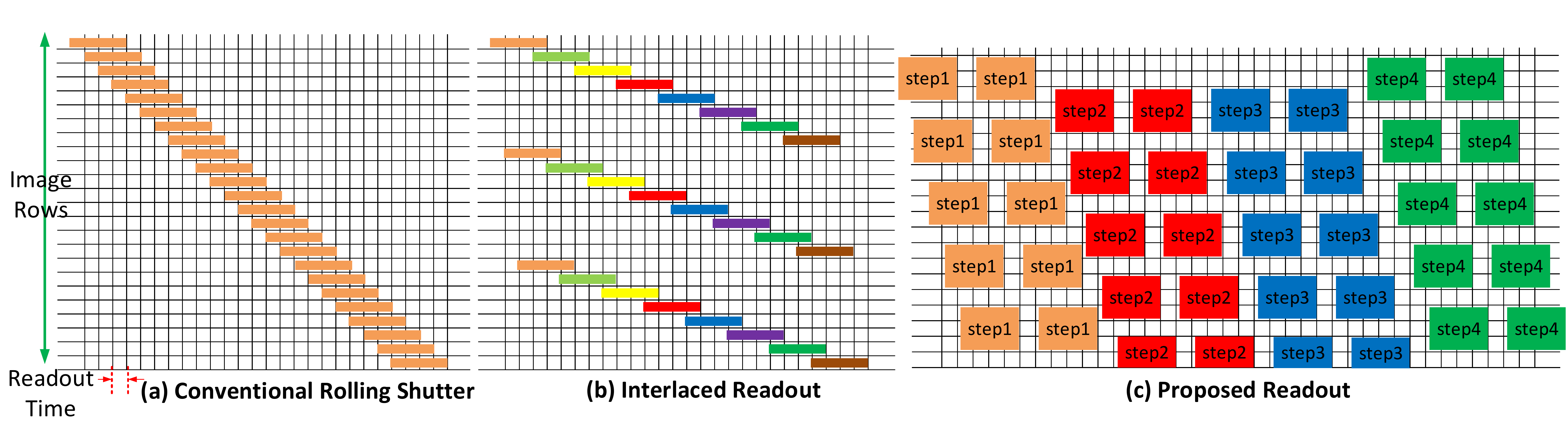}
\caption{The exposure and readout sequence diagram of (a) conventional Rolling Shutter, (b) Interlaced Readout, (c) Proposed Readout.}
\label{interlace}
\end{figure*}

The pixel units in the same column are connected to the same column readout circuit. Each tile includes 3 column readout circuits. To read out the computing results of every three rows of tiles simultaneously, pixel units with the row number x (x = 4n + 3, n = 0, 1, 2, 3...) are connected to three independent row enabling signals $C_1$, $C_2$, $C_3$, as shown in Fig. \ref{wiring}(b). As shown in Fig. \ref{conv33}, the bit lines $C_1$, $C_2$, and $C_3$ output the computing results of the three consecutive rows of tiles in the vertical direction. In this way, the tiles in one group are read out simultaneously.

The processing sequence of the convolution operation is shown in Fig. \ref{array convolution}. The subscript n represents the $n^{th}$ group. 
As shown in Fig. \ref{array convolution}(b), a tile is reset immediately after each readout operation and then begins the exposure for the next readout. Assuming the resolution is 128$\times$128, the convolution kernel size is 3$\times$3, the stride is 2, and each step contains 32 rows of convolution kernel results, then each step only needs 11 readout operations. The whole process of array convolution operation is described in Algorithm \ref{alg1}.

This processing sequence has an exception case. The link between pixel units in the same row are realized by splicing transistors controlled in the column direction. As shown in Fig. \ref{cis-method}(b), the splicing transistors in the column direction have to be switched when moving from step 2 to step 3 (also when moving from step 4 to step 1). The exposure for step 3 cannot be started until the step 2 has been finished completely. A waiting time should be added between the two steps. For each convolution kernel, this increases the number of equivalent exposure times required from eight to ten. This influence can be ignored when the illumination is strong enough and the ADC frequency becomes the main factor to determine the maximum frame rate.

The readout time of an entire step is $(n_{rd}-1) T_{rd}$, where $T_{rd}$ is the readout time for each tile and $n_{rd}$ is the number of the readout operations in each step. If the reset interval is $T_{rst}$, and the maximum exposure time is $T_{expo}$, the reset and exposure stages need to be finished before the next readout operation,

\begin{equation}
(n_{rd}-1) T_{rd} > T_{rst} + T_{expo}
\label{times relationship}
\end{equation}

As shown in Fig. \ref{interlace}, the exposure and readout sequence of the proposed architecture is similar to Interlaced Readout method\cite{interlaced}, which divides the readout time for one frame into K sub-images. Compared with the conventional Rolling Shutter as shown in Fig. \ref{interlace} (a), the Interlaced Readout in Fig. \ref{interlace} (b) reduces the skew and the time lag in these sub-images K times to support high speed photography, at the cost of the reduction of vertical spatial resolution. Our proposed architecture divides the computing of one frame into 8 groups (each step exposures 2 times for the processing of positive or negative weights) as shown in Fig. \ref{interlace} (c), which improves the readout speed and resource utilization.

\begin{algorithm}[t]
	\setstretch{1.1}
	\caption{Array Convolution Operation} 
	\label{alg1} 
	\begin{algorithmic}[1]
			\For{each step (1-4)}
				\State Divide the array into independent tiles.
				\For{positive and negative weights(+,-)}
					\For{$j=1$ to $n_{rd}$ (the group serial number)}
					\State Read the results of the three rows of tiles 
					\Statex \qquad\qquad\ via three column readout circuits, respectively.
					\State Reset and exposure group $j$ for next readout. 
					\Statex \qquad\qquad\ (While continuing to read the next groups 
					\Statex \qquad\qquad\ $j+1$, $j+2$...) 
					\EndFor
				\State Calculate the total results by eq. (3).
				\EndFor
			\EndFor
	\end{algorithmic} 
\end{algorithm}

\subsection{Implementation of Different Convolution Kernel Size}

\begin{figure}[!t]
\centering
\includegraphics[width=0.5\textwidth]{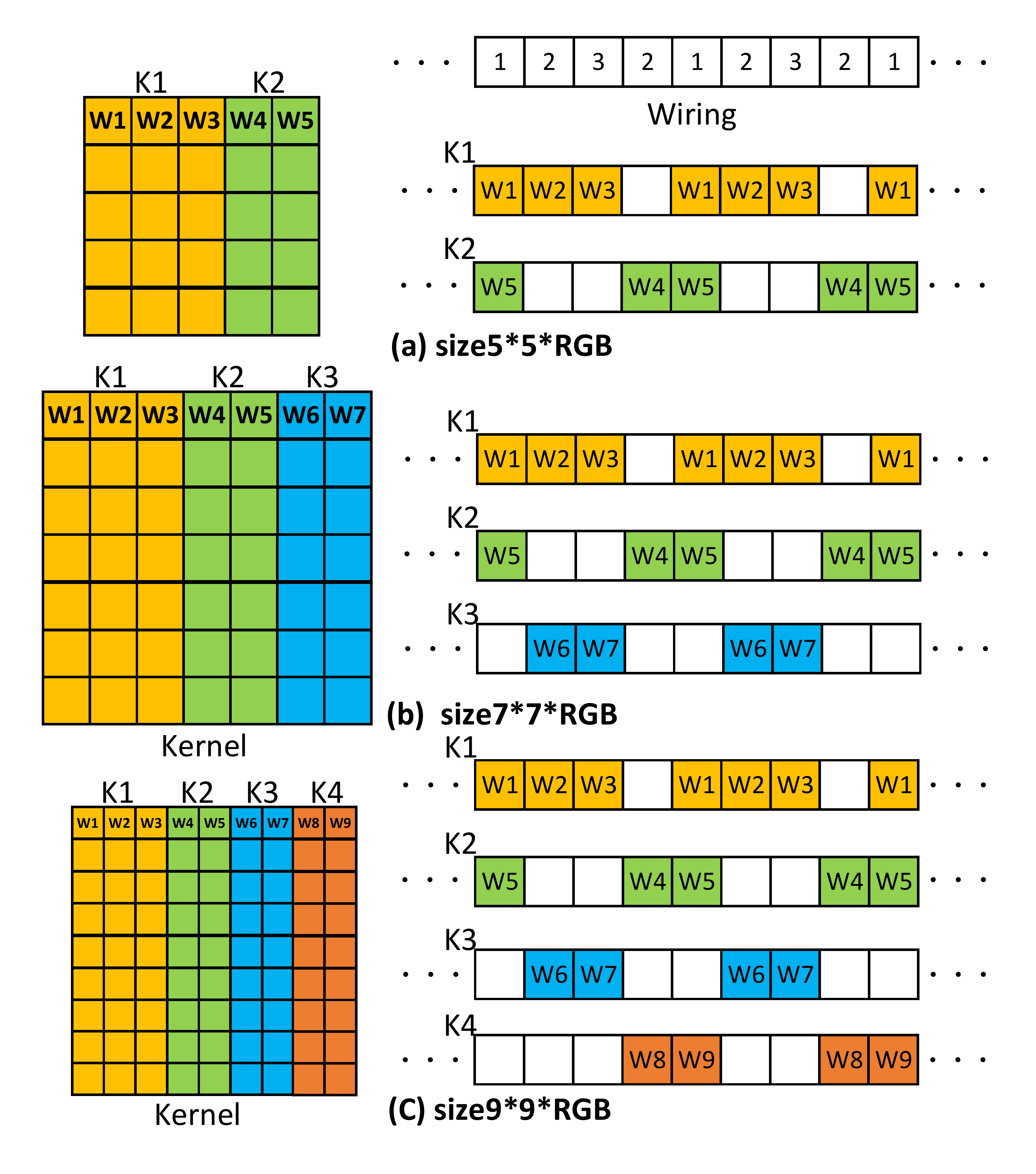}
\caption{The convolution implementations of kernel splitting. The convolution kernel size is (a) 5$\times$5, (b) 7$\times$7, and (c) 9$\times$9.}
\label{split mode}
\end{figure}

We propose a ``kernel splicing'' method to support different kernel sizes with the same wiring method. As shown in Fig. \ref{split mode}(a), 5$\times$5 convolution operation can be realized by two 5$\times$3 convolution operations. The 5$\times$3 convolution operation is similar to the 3$\times$3 convolution operation. The main difference is that the convlink wires link 5$\times$3 pixel units together in the 5$\times$3 convolution operation. The number of steps is changed to 6 instead of 4 because three steps are required in the column direction to avoid overlapping. Each group still has 3 rows of tiles during readout operation. Another two examples for 7$\times$7 and 9$\times$9 kernel sizes are shown in Fig. \ref{split mode}(b) and (c), respectively. In this way, different kernel sizes can be realized by using the same hardware circuits. 

Assuming the kernel size is r$\times$r and the stride is $s$, the total number of steps is $\frac{r+1}{s}(r-1)$, where the ratio $\frac{r+1}{s}$ needs to round up to an integer if necessary. The equivalent exposure times for each channel are $[\frac{2(r+1)}{s}+1](r-1)$. For a fixed height of the pixel array H (128 in our case), the total number of output rows in each step is $\frac{H}{r + 1}$. Since each readout operation contains three output rows, the minimum ADC conversion rate can be calculated by

\begin{equation}
f_{ADC(min)}=\frac{2fnH(r-1)}{3s}
\label{f-adc}
\end{equation}
where $f$ is the frame rate and $n$ is the number of channels. The minimum conversion rate of ADC is proportional to the frame rate $f$, the channel number $n$, and the kernel size $r$. It is inversely proportional to the stride $s$.

The actual frame rate $f_{real}$ is defined as the product of frame rate and the output channel number $f\times n$. With a fixed maximum exposure time $T_{expo}$, the maximum frame rate can be calculated by

\begin{equation}
f_{real(max)}=\frac{s}{[2(r+1)+s](r-1)T_{expo}}
\label{f-real}
\end{equation}

The maximum frame rates and the minimum ADC conversion rates at different kernel sizes are tabulated in Table \ref{method}. For example, the resolution is 128$\times$128, the stride is 2, the kernel size is 3$\times$3, the stride is 2, and the output channel number is 64, the maximum exposure time is 26.04 us. $f_{ADC(min)}$ is based on $f_{real(max)}$ in each condition. When kernel size increases, the conversion rate of ADC will decrease because both actual frame rate and readout operation frequency decrease. As a result, the maximum frame rate decreases. 

\begin{table}
\renewcommand{\arraystretch}{1.3}
\caption{Rate calculation with different kernel sizes.}
\label{method}
\centering
\begin{tabular}{|c|c|c|}
\hline
\textbf{operation} & \textbf{minimum ADC} & \textbf{maximum real}\\
\textbf{condition} & \textbf{conversion rate} & \textbf{frame rate}\\
\hline
3$\times$3 & 327.68 KHz & 3840\\
\hline
5$\times$5 (splicing) & 234.06 KHz & 1371\\
\hline
7$\times$7 (splicing) & 182.04 KHz & 711\\
\hline
9$\times$9 (splicing) & 148.95 KHz & 436\\
\hline
\end{tabular}
\end{table}

%
%

\section{Simulation Results}
\begin{figure}[!t]
\centering
\includegraphics[width=0.48\textwidth]{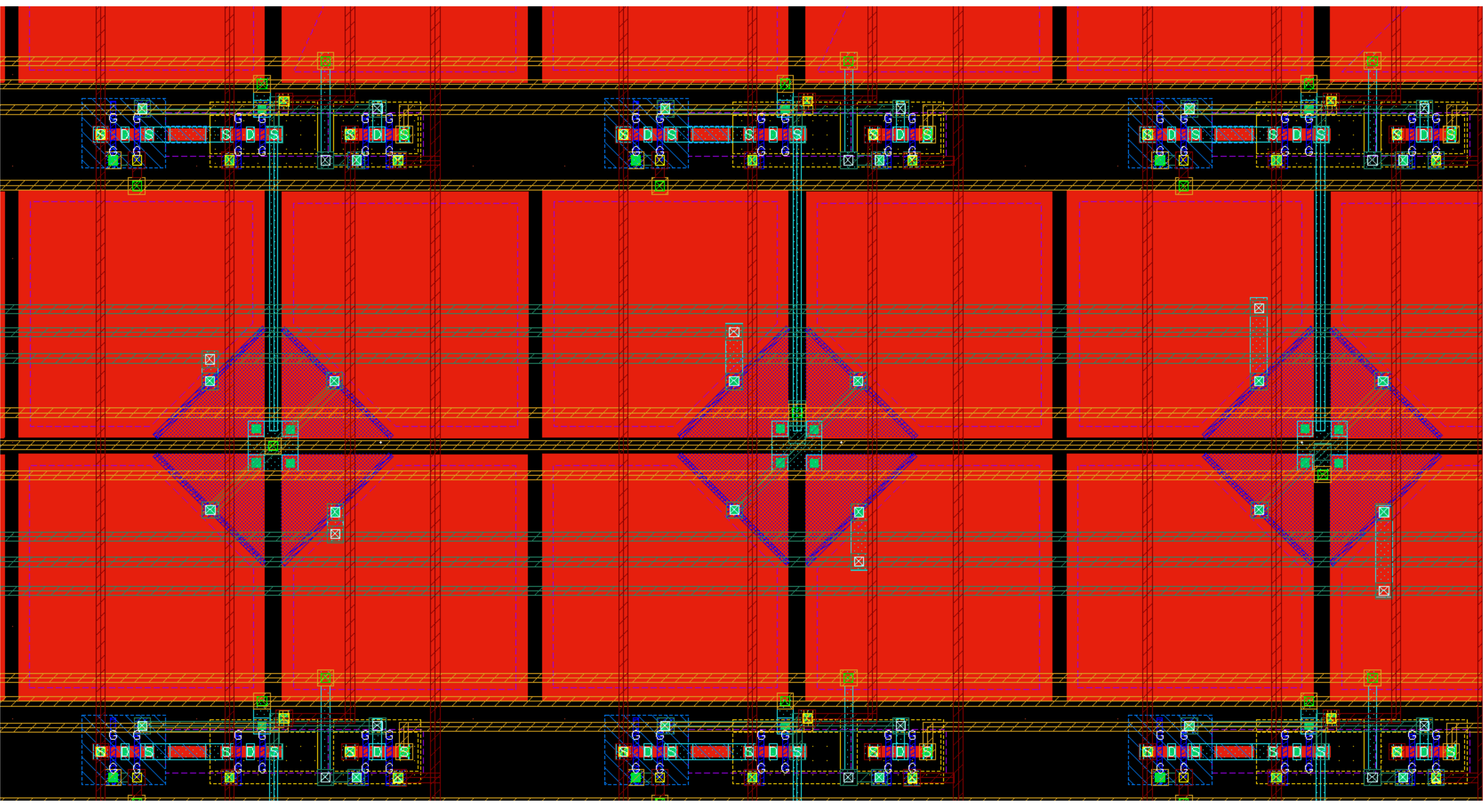}
\caption{The layout design of the proposed pixel.}
\label{layout}
\end{figure}

Our proposed architecture was implemented with TowerJazz 0.18 um CMOS process. The circuits are simulated with Cadence Virtuoso and Spectre. An analytic model taken from \cite{PDmodel} is used to simulate the response of the photodiode. The capacitance of the shared FD node in each pixel units is 22.2 fF, and the responsivity $R_{\lambda}$ is about 0.35 A/W at $\lambda$ = 555 nm. The photodiode size is set to 10 um$\times$ 10 um to collect enough photons. The photocurrent can be estimated as

\begin{equation}
I_{ph} = R_{\lambda} (P_{in}\times area) 
\label{frame-cal}
\end{equation}
where $P_{in}$ represents the input optical power in $W/m^2$. The layout design of the pixel is shown in Fig. \ref{layout}. 
\begin{figure}[!t]
\centering
\includegraphics[width=0.45\textwidth]{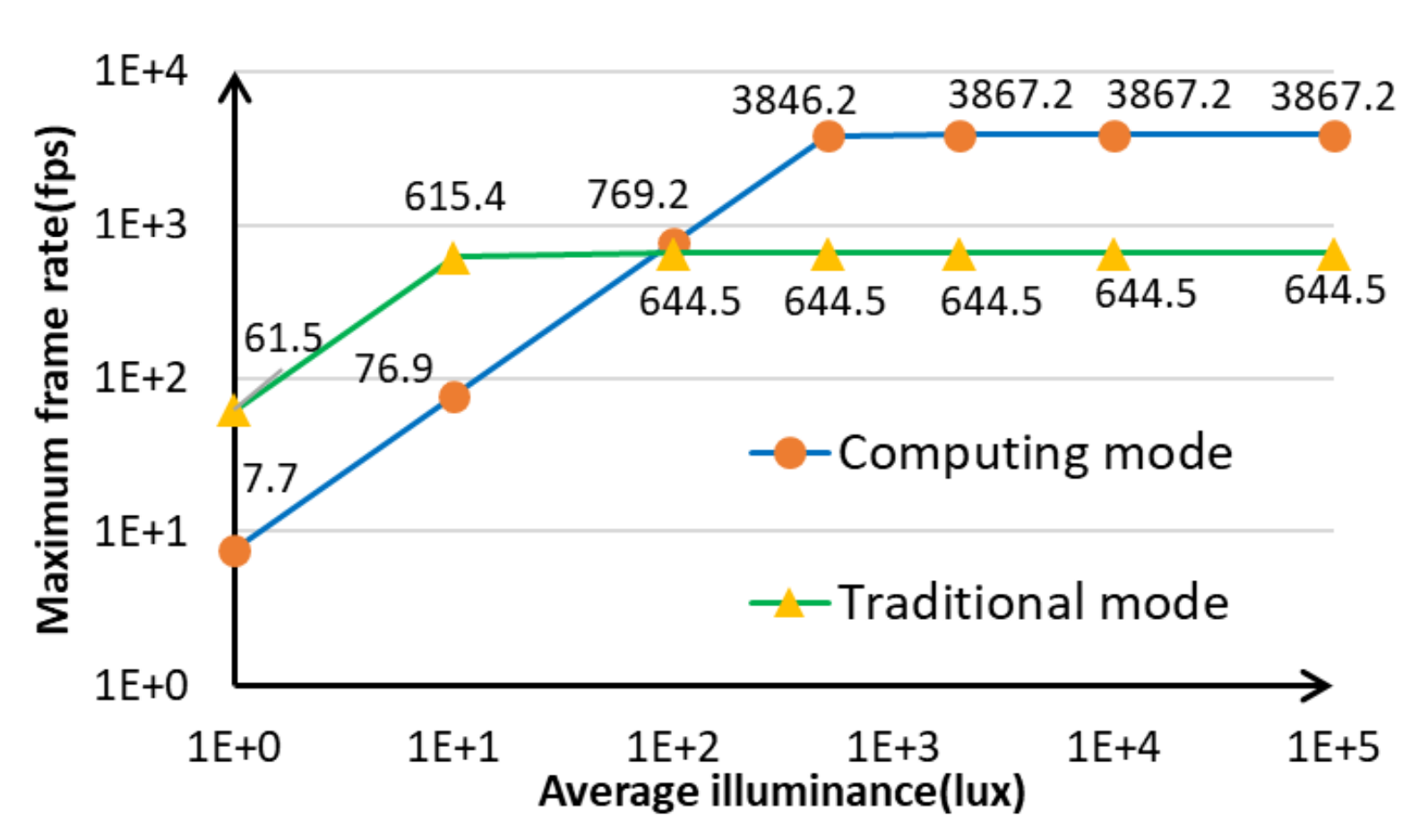}
\caption{The maximum frame rate of proposed CIS working in Computing mode and Traditional mode under different illumination conditions.}
\label{frame}
\end{figure}

\begin{figure}[!t]
\centering
\includegraphics[width=0.45\textwidth]{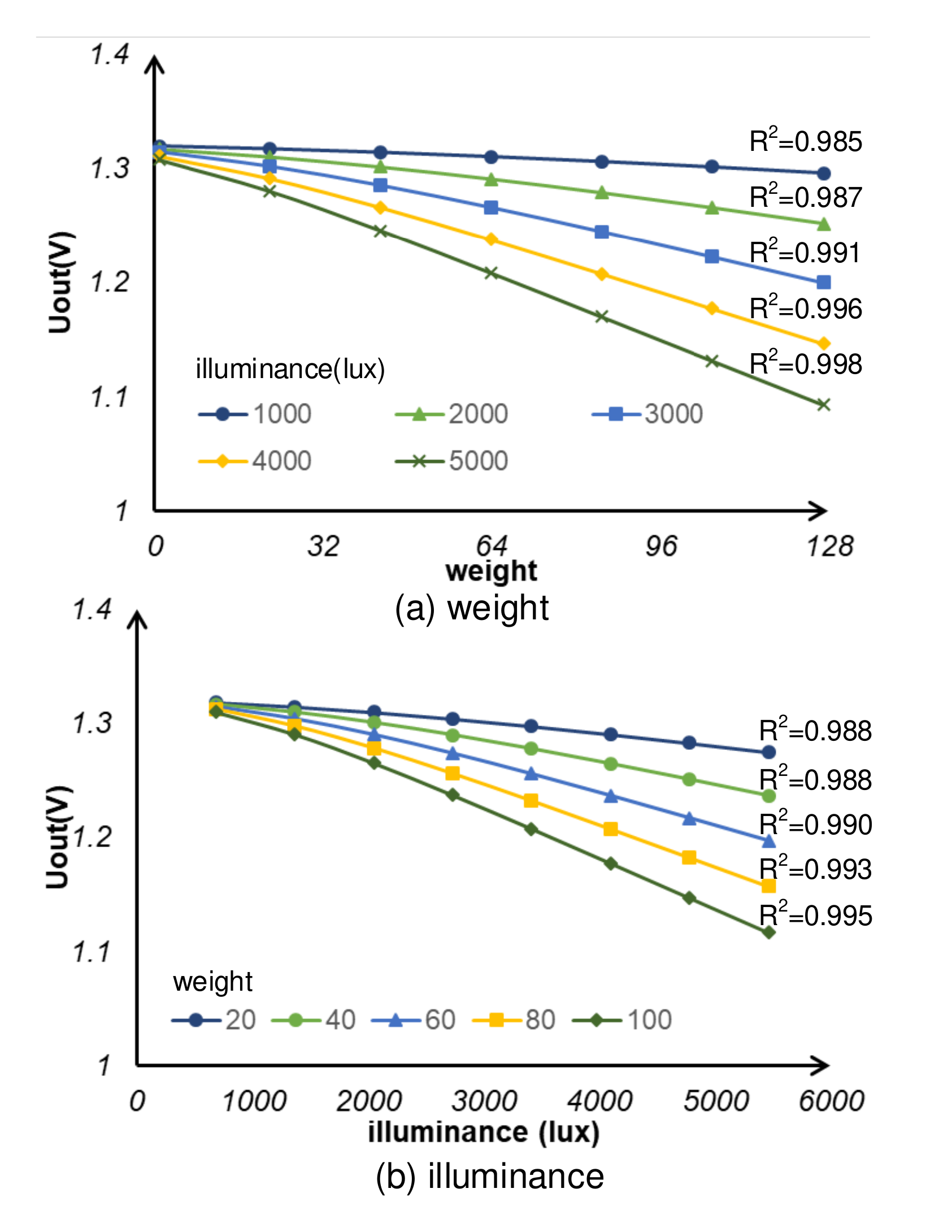}
\caption{Simulation of MAC operations. (a) The relationship between readout voltage and weight values at various illuminance. (b) The relationship between readout voltage and illuminance at various weight values.}
\label{sim-mac}
\end{figure}

The maximum frame rate under different illumination conditions of proposed CIS working in Computing mode and Traditional mode are shown in Fig. \ref{frame}. In this simulation, the array size is 128$\times$128, kernel size is 3$\times$3, the stride is 2, and the ADC conversion rate is 330 KHz. When the light intensity is low, the exposure time is the main factor to determine the maximum frame rate. When the light intensity is high enough, the ADC frequency limits the maximum frame rate. The frame rate of the proposed CIS in Traditional mode is the same as a conventional 1.75T CIS. The proposed Computing mode requires eight times the exposure time of conventional 1.75T CIS for the convolution operation but reduces the readout density. With strong illumination, our proposed CIS in Computing mode outperforms the conventional 1.75T CIS because of the lower readout density.

\subsection{Linearity Simulation}

Although FD node behaves electrically as a capacitor, it is quite different from a real capacitor. That is, the non-idealities of the FD node need to be considered in the simulation. Thus, we use a resistor (R = 8.07e15 $\Omega$, estimated based on the data in \cite{leakeage}) connected in parallel with the FD node to simulate the leakage of the FD node.

Fig. \ref{sim-mac} (a) and (b) show the readout voltage $U_{out}$ under different weight and illuminance, respectively. Linear fitting results of both figures show that the $R^2$ are all above 0.98, indicating high linearity and accuracy of the proposed architecture. 

\begin{table*}[!t]
\renewcommand{\arraystretch}{1.2}
\caption{Power consumption analysis.}
\label{tab-performance}
\centering
\begin{tabular}{|m{4cm}<{\centering}|m{1.85cm}<{\centering}|m{1.85cm}<{\centering}|m{1.85cm}<{\centering}|m{1.85cm}<{\centering}|m{1.6cm}<{\centering}|m{2.2cm}<{\centering}|}
\hline
\textbf{condition} & \textbf{Pixel Power (uW)}&\textbf{Readout Power (uW)}& \textbf{ADC Power (uW)}& \textbf{Total Power (uW)} & \textbf{Efficiency (TOPS/W)} & \textbf{FoM (pJ/pixel/frame)}\\
\hline
60FPS, 3$\times$3, s=2 &63.94&4.02&177.17& 245.13 & 4.62 & 3.90\\
\hline
120FPS, 3$\times$3, s=2 &127.87&8.03&354.33& 490.25 & 4.62 & 3.90\\
\hline
60FPS, 5$\times$5 (splicing), s=2 &177.60&4.02&177.17& 358.79 & 8.77 & 5.70\\
\hline
60FPS, 5$\times$5 (splicing), s=4 &44.40&1.01&44.29& 89.70 & 8.77 & 1.43\\
\hline
60FPS, 7$\times$7 (splicing), s=2 &348.10&4.02&177.17& 529.29 & 11.65 & 8.41\\
\hline
60FPS, 7$\times$7 (splicing), s=4 &87.02&1.01&44.29& 132.32 & 11.65 & 2.10\\
\hline
\end{tabular}
\end{table*}

\begin{table*}[!t]
\renewcommand{\arraystretch}{1.2}
\caption{Performance Comparison}
\label{tab-compare}
\centering
\begin{threeparttable}
\begin{tabular}{|m{1.9cm}<{\centering}|m{2.2cm}<{\centering}|m{2cm}<{\centering}|m{1.8cm}<{\centering}|m{1.8cm}<{\centering}|m{2cm}<{\centering}|m{2cm}<{\centering}|m{1.8cm}<{\centering}|}
\hline
 & \textbf{2017ISSCC\cite{3dstacked}} & \textbf{2019ASSCC\cite{2019ASSC}} & \textbf{2020TCASii\cite{2020TCASii}} & \textbf{2020TCASi\cite{nearsensor}} & \textbf{2021ISSCC\cite{2021pip}} & \textbf{this work}\\
\hline
\textbf{Process} &90 nm/60 nm & 180 nm & 180 nm & 180 nm & 65nm & 180 nm\\
\hline
\textbf{Supply} & 3.3/2.9/1.8/1 V & 0.5 V & 1.8 V & 2 V & 0.8 - 1.2 V & 1.8 V\\
\hline
\textbf{Array Size} & 1296$\times$976 & 128$\times$128 & 32$\times$32 & 32$\times$32 & 160$\times$128 & 128$\times$128\\
\hline
\textbf{Pixel Size (um$^2$)} & 3.5$\times$3.5 & 7.6$\times$7.6 & 110$\times$110 & 40$\times$40 & 9$\times$9 & 30$\times$30\\
\hline
\textbf{Frame Rate} & 1000 fps & 480 fps & 1000 fps & 100 fps & 96 - 1072 fps & 3840 fps\\
\hline
\textbf{data/weight width} & 4 bit/NA & 8 bit/4 bit & 1 bit/1 bit & 1 bit/1 bit & 8 bit/1.5 bit & 8 bit/8 bit\\
\hline
\textbf{Feature} & Spatial-Temporal processing & 1st-layer CNN, ED, Blur, Sharpen & 1st-layer BNN & 1st-layer BNN &2 - 64 conv, ROI & 1st-layer CNN\\
\hline
\textbf{Processing} & Digital & Analog & Analog & Analog & Analog & Analog\\
\hline
\textbf{Memory} & Yes(Digital) & No & No & Yes(Analog) & No & No\\
\hline
\textbf{Pixel complexity} & 3T & 4T & 88T, 32 caps & 3T & 40T, 1 cap & 2.5T\\
\hline
\textbf{Method} & 3D-stacked & switching-current-and-integration & dedicated SRAM &Kernel-Readout &pulse modulation& pulse modulation\\
\hline
\textbf{Power} & 363 mW  & 91 uW & 12.16 uW &1.8 mW & 42 - 206 uW & 89.7 - 529.29 uW \\
\hline
\textbf{Efficiency (TOPS/W))} & 0.77 & 1.51 & 2.64 & 1.09 & 0.15-3.64 &\textbf{4.62 - 11.65}\\
\hline
\textbf{FoM (pJ/pixel/fr)} & 286.98 & 11.65 & 11.88 & NA & 2.5-103.9 & 1.43 - 8.41\\
\hline
\end{tabular}
\end{threeparttable}
\end{table*}

\begin{figure*}[!t]
\centering
\includegraphics[width=0.9\textwidth]{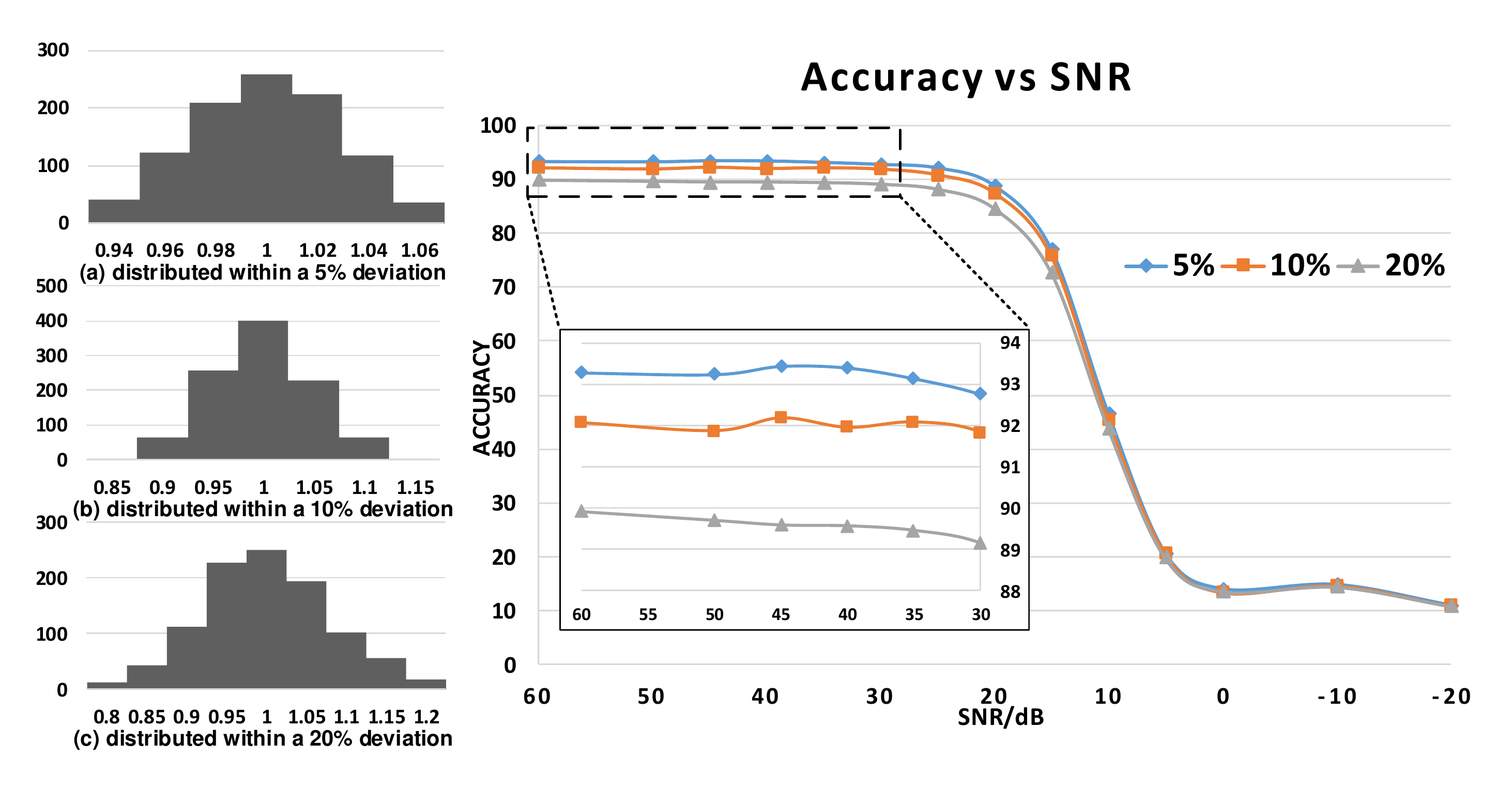}
\caption{Relationship between CNN accuracy and three types of disturbance.}
\label{sim-cnn}
\end{figure*}

\subsection{Performance Analysis}
The power consumption and performance comparison under different conditions are tabulated in Table \ref{tab-performance}. The illumination is about 1500 lux. The array size for all situations is 128$\times$128, and the number of output channels is 64. The computing results with 5$\times$5 and 7$\times$7 kernels are estimated based on the kernel splicing. The FoM (pJ/pixel/frame) presents the energy consumed per frame (each channel counts once) per pixel. The computing efficiency (TOPS/W) shows the computing amount divided by power consumption. The total computing amount is calculated as the product of the size of the output array, the number of input channels, the number of output channels, the frame rate, and the computing amount for one kernel. When the kernel size is 3$\times$3, each kernel needs 18 OPs, and the total amount is $64 * 64 * 4 * 64 * 60 * 18 = 1132462080$ OPs. The power consumption of ADCs is calculated according to the data in \cite{adc18}. Advanced CMOS technology will significantly reduce the power consumption of ADC.

Table \ref{tab-performance} shows that the power consumption of our proposed sensor circuit is determined by the frame rate and the convolution kernel size. When the convolution kernel size keeps constant, the power consumption increases with the frame rate. With the same frame rate, the larger convolution kernel size leads to higher power consumption.

Computational efficiency (TOPS/W) changes remarkably in different conditions. The power consumption mainly comes from three parts: the convolution operation, the readout circuit, and the column ADCs. Though the number of convolution operations and the number of readouts vary in the same proportion, the cost of ADCs remains unchanged, leading to increased computational efficiency when the amount of computation increases. For example, the efficiency is 4.62 TOPS/W and the frame rate is 60 fps when the kernel size is 3$\times$3 and the stride is 2. It is increased to 11.65 TOPS/W when the kernel size is 7$\times$7.

FoM (pJ/pixel/frame) increases with the amount of computation but decreases with frame rate because of the same reason as the computational efficiency. It mainly represents the influence of convolution kernel size on power consumption. The increase in convolution kernel size will lead to a rise in the amount of computation and power consumption.

The change of stride will lead to different amount of computation. If the stride is doubled, the amount of computation and readout times will be reduced to a quarter, which leads to a decrease in power consumption and FoM.

As shown in Table \ref{tab-compare}, the computing efficiency of our proposed architecture is up to 11.65 TOPS/W, which is three times as high as the latest in-sensor computing schemes. Since our proposed architecture integrates MAC operation with the pixel exposure, no additional analog computing circuit is used, leading to this high efficiency. 
The energy consumption of a single frame of figure has been estimated when a 7$\times$7 kernel is adopted and stride is 2.
A traditional CIS + DLA vision system consumes about 133.74 uJ (based on \cite{ulvpwm,DLA}) and the proposed scheme costs only 8.82 uJ, which shows a 15 times reduction of the energy consumption. 
The energy consumption of the CIS is estimated based on the scaling of resolution and frame rate, and the energy consumption of the DLA is estimated by the product of the computational efficiency of the DLA and the computing amount. 


\begin{figure*}[!t]
\centering
\includegraphics[width=0.8\textwidth]{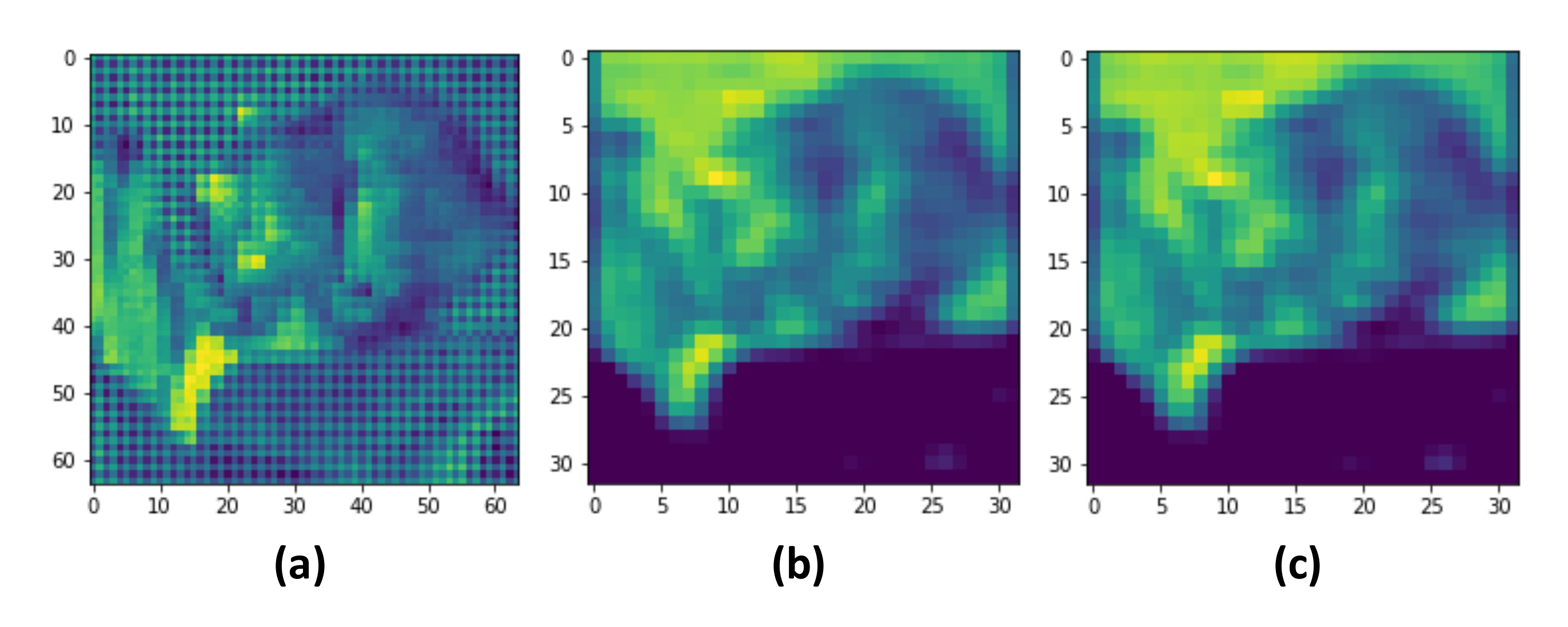}
\caption{(a) An input test figure (with RGGB pattern). (b) A channel of the output feature map of the 1-st layer CNN performed by the proposed CIS and (c) by CPU.}
\label{cis-feature}
\end{figure*}
\subsection{Noise Analysis}
Operations in the analog domain are affected by undesirable factors such as noise and variations. In this section, we analyze the robustness of our proposed circuits.

The temporal noise and Fixed Pattern Noise (FPN) are the most fundamental nonideality in CIS\cite{CIS2005}. The sample noise at the end of integration can be expressed as the sum of: (1) integrated shot noise $Q_{shot}$, (2) reset noise $Q_{reset}$, (3) readout circuit noise $Q_{read}$ due to readout device thermal and flicker (or 1/f) noise, (4) offset FPN due to device mismatches $Q_{FPN}$, (5) offset FPN due to dark current variation, commonly referred to as dark signal nonuniformity (DSNU), and (6) gain FPN, commonly referred to as PRNU. For a conventional CIS with a CDS operation, the total noise can be estimated by
\begin{equation}
\begin{split}
S_{2} - S_1 = Q_{shot} - Q_{1,read}
+ Q_{2,read} \\+ Q_{DSNU} + Q_{PRNU}
\end{split}
\label{rob-cis}
\end{equation}
in which the reset noise and the offset FPN are suppressed but the read noise power is increased.

In our proposed scheme, the CIS performs two exposures and readouts for positive and negative weights, respectively. After the subtract operation by the digital circuits, the noise can be estimated by
\begin{equation}
\begin{split}
\\S_p - S_n = (Q_{p,shot} - Q_{n,shot}) + (Q_{p,reset} - Q_{n,reset}) \\
+ (Q_{p,read} - Q_{n,read})+ (Q_{p,PRNU} - Q_{n,PRNU})
\end{split}
\label{rob-pip}
\end{equation}
in which the offset FPN are suppressed, and most of low frequency noise will be reduced due to the short time interval between two readout (less than 32 us, about 30 KHz). Therefore, the proposed CIS does not add additional noise. In addition, the MAC operation on the pixel array may benefit the SNR, in the case of a 3$\times$3 kernel this can be explained as

\begin{equation}
noise = E[(\frac{\sum n_{all}}{9})^2] = \frac{1}{9^2}D(\sum n_{all})= \frac{\sigma ^2}{9}
\label{rob-cal-noise}
\end{equation}
\begin{equation}
SNR = \frac{power}{noise} = 9\frac{power}{\sigma ^2}
\label{rob-cal-snr}
\end{equation}
in which $n_{all}$ represents the total noise.

\subsection{Algorithm Robustness}
Through network simulation with Cifar-10\cite{CIFAR10} dataset and Resnet-18, the accuracy of CNN changes with SNR or mismatch as shown in Fig. \ref{sim-cnn}. As the proposed CIS only supports 1st-layer CNN, the rest of the computation is performed externally. Three normal distributions with different variances of the equivalent capacitors are used to simulate the mismatch, and the distributions are shown in Fig. \ref{sim-cnn}. The results only have a negligible accuracy loss when SNR is more than 40 dB. The typical SNR value for CIS is 40 dB - 60 dB \cite{2020TCASii}. Fig. \ref{cis-feature} shows a comparison of the input RGGB figure and the output feature map of the 1-st layer CNN performed by CPU or the proposed CIS. The mean RMS error for the 1-st layer output of the proposed CIS is shown in Table. \ref{rms-error}. As expected with the reduction in the SNR, the RMS error increases.

Because the first layer of CNN is sensitive to pruning and quantization, the conventional network model compression algorithms often leave the first layer untouched. For example, most layers may perform a 4-bit quantization, but the 1-st layer needs at least 8-bit to keep accuracy. In this case, DLAs must support 8-bit computing to support the entire net, which leads to extra costs and reduction of system efficiency. Moreover, due to the small number of input channels, DLAs' PEs are often not fully utilized for the first layer. Therefore, this design can improve the computational efficiency of the subsequent DLAs, leading to higher performance of the entire machine vision system.

\begin{table}
\renewcommand{\arraystretch}{1.2}
\caption{Mean RMS error for the 1-st layer output in different condition}
\label{rms-error}
\centering
\begin{tabular}{|c|c|c|c|c|}
\hline
\textbf{SNR} & 60dB & 40dB & 20dB & 0dB\\
\hline
5\% deviation& 6.78e-3&7.12e-3&2.25e-2&1.77e-1\\
\hline
10\% deviation & 1.13e-2&1.15e-2&2.43e-2&1.77e-1\\
\hline
20\% deviation& 1.80e-2&1.82e-2&2.80e-2&1.78e-1\\
\hline
\end{tabular}
\end{table}

\subsection{Performance under Other Resolutions}
In this subsection, we discuss the power consumption, frame rate, and Efficiency of the architecture under different resolutions to show the scalability.

\begin{figure}[!t]
\centering
\includegraphics[width=0.45\textwidth]{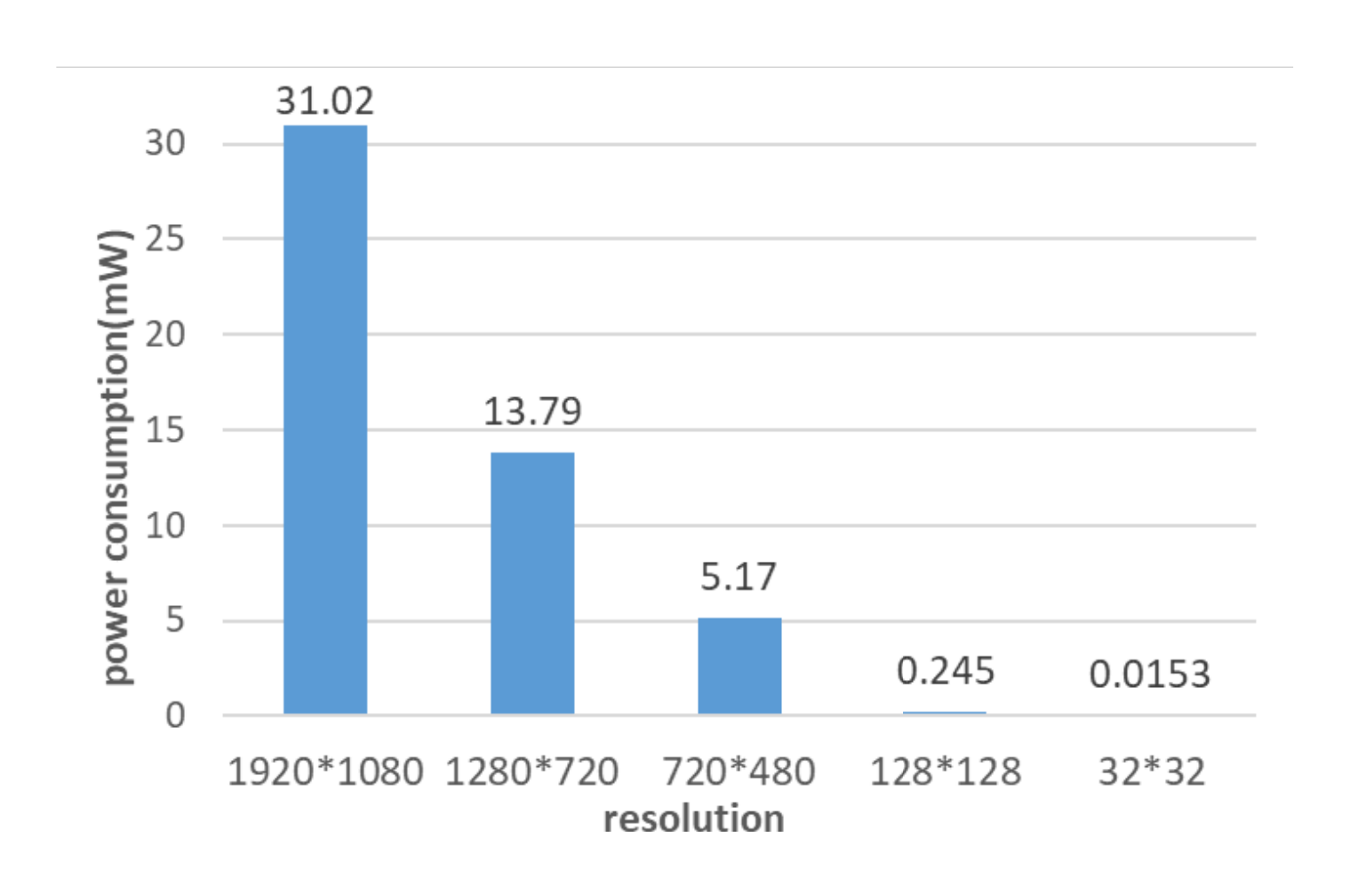}
\caption{The power consumption under different resolution. The illumination is about 1500 lux.}
\label{resolution}
\end{figure}

\subsubsection{Power Consumption}
Fig. \ref{resolution} shows the power consumption of the proposed architecture under different resolutions. The power consumption is basically proportional to the amount of calculation thus increases with the resolution. Thus, the computation efficiency is independent of the resolution.

\subsubsection{Maximum Frame Rate}
According to Eq. \ref{f-real}, the maximum frame rate limited by exposure time is independent of resolution. As shown in Eq. \ref{f-adc}, the minimum ADC frequency is proportional to the height of the array. The minimum ADC frequency under other resolutions is tabulated in Table \ref{adc-resolution}. The operation condition is all set to 3$\times$3 kernel and stride is 2. 

\begin{table}
\renewcommand{\arraystretch}{1.2}
\caption{Rate calculation under different resolutions.}
\label{adc-resolution}
\centering
\begin{tabular}{|c|c|}
\hline
\textbf{resolution} & \textbf{minimum ADC frequency}\\
\hline
1920$\times$1080 & 2.76 MHz\\
\hline
1280$\times$720 & 1.84 MHz\\
\hline
720$\times$480 & 1.23 MHz\\
\hline
128$\times$128 & 327.68 KHz\\
\hline
32$\times$32 & 81.92 KHz\\
\hline
\end{tabular}
\end{table}

\section{Conclusion}
In this work, a PIP architecture has been proposed to perform the first layer convolution operation of CNN. It supports a variety of different convolution kernel sizes and parameters. The simulation results have shown that our proposed scheme functions correctly with good linearity. When the convolution kernel is 7$\times$7, the step size is 2, and the channel number is 64 at 60 fps, the proposed architecture consumes 529.29 uW power and has a computational efficiency up to 11.65 TOPS/W. It is suitable for application scenarios with tight requirements on power consumption, such as daily monitoring and Internet of Things (IoT) devices.

\ifCLASSOPTIONcaptionsoff
  \newpage
\fi

\bibliographystyle{IEEEtran}
\bibliography{IEEEabrv,mylib}

\begin{IEEEbiography}[{\includegraphics[width=1in,height=1.25in,clip,keepaspectratio]{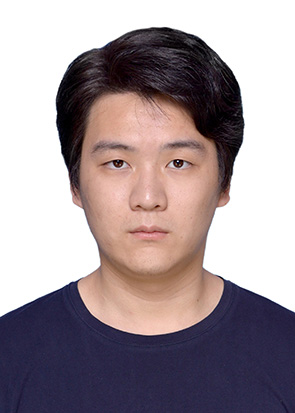}}]{Ruibing Song}
(Student Member, IEEE) received a bachelor’s degree from the College of Electrical Engineering, Zhejiang University, in 2020. He is currently pursuing a master’s degree at the College of Information Science \& Electronic Engineering, Zhejiang University. He is interested in in-sensor computing and in-memory computing.
\end{IEEEbiography}

\begin{IEEEbiography}[{\includegraphics[width=1in,height=1.25in,clip,keepaspectratio]{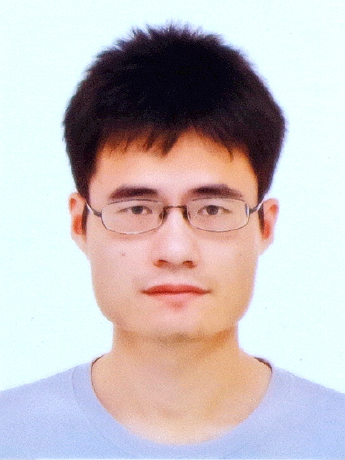}}]{Kejie Huang}
(Senior Member, IEEE) received the Ph.D. degree from the Department of Electrical Engineering, National University of Singapore (NUS), Singapore, in 2014. He has been a Principal Investigator with the College of Information Science Electronic Engineering, Zhejiang University (ZJU), since 2016. Before joining ZJU, he has spent five years in the IC design industry, including Samsung and Xilinx, two years in the Data Storage Institute, Agency for Science Technology and Research (A*STAR), and another three years in Singapore University of Technology and Design (SUTD), Singapore. He has authored or coauthored more than 40 scientific articles in international peer-reviewed journals and conference proceedings. He holds four granted international patents, and another eight pending ones. His research interests include low power circuits and systems design using emerging non-volatile memories, architecture and circuit optimization for reconfigurable computing systems and neuromorphic systems, machine learning, and deep learning chip design. He currently serves as the Associate Editor of the IEEE TRANSACTIONS ON CIRCUITS AND SYSTEMS-PART II: EXPRESS BRIEFS.
\end{IEEEbiography}

\begin{IEEEbiography}[{\includegraphics[width=1in,height=1.25in,clip,keepaspectratio]{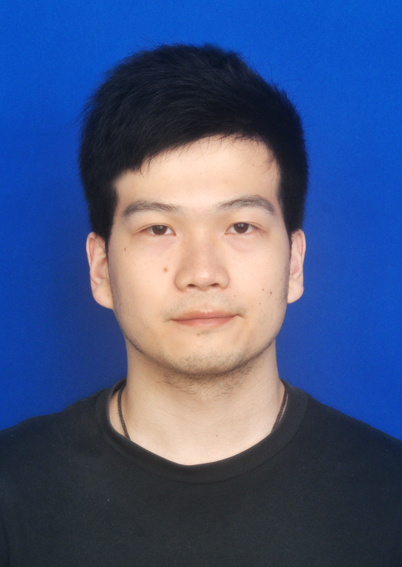}}]{Zongsheng Wang}
(Student Member, IEEE) received a bachelor’s degree from the College of Electrical Engineering, Zhejiang University, in 2020. He is currently pursuing a master’s degree at the College of Information Science \& Electronic Engineering, Zhejiang University. He is interested in in-sensor computing, low power digital circuit design and deep learning accelerator.
\end{IEEEbiography}

\begin{IEEEbiography}[{\includegraphics[width=1in,height=1.25in,clip,keepaspectratio]{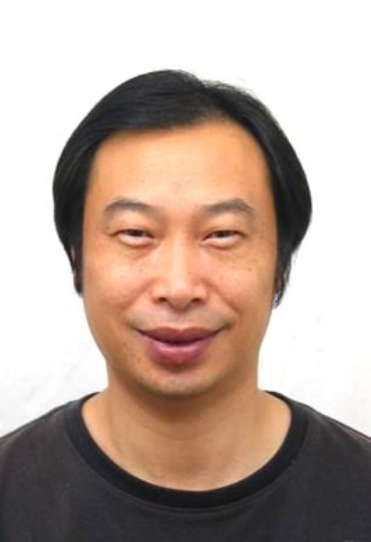}}]{Haibin Shen}
is currently a Professor with Zhejiang University, a member of the second level of 151 talents project of Zhejiang Province, and a member of the Key Team of Zhejiang Science and Technology Innovation. His research interests include learning algorithm, processor architecture, and modeling. His research achievement has been used by many authority organizations. He has published more than 100 papers in academic journals, and he has been granted more than 30 patents of invention. He was a recipient of the First Prize of Electronic Information Science and Technology Award from the Chinese Institute of Electronics, and has won a second prize at the provincial level.
\end{IEEEbiography}

\end{document}